\documentclass[12pt]{article}
\usepackage{amsmath}
\usepackage{amssymb}
\usepackage{latexsym}
\usepackage{epsfig}
\usepackage{graphicx}
\usepackage{slashed}

\usepackage{amssymb}
\usepackage{euscript}

\catcode`\@=11 \@addtoreset{equation}{section}

\def\0{\nonumber}

\newcommand{\ben}{\begin{eqnarray}\displaystyle}
\newcommand{\een}{\end{eqnarray}}

\newcommand\ee{\end{eqnarray}} 
\newcommand\be{\begin{eqnarray}}
\newcommand\ba{\begin{array}} 
\newcommand\ea{\end{array}}
\newcommand\eeq{\end{equation}} 
\newcommand\beq{\begin{equation}}
\newtheorem{HW}{Problem}[section]

\newcommand{\dsl}{\raise.15ex\hbox{/}\kern-.57em\partial}
\def\aslash{\raise.15ex\hbox{/}\mkern-14mu A}
\newcommand{\pslash}{\raise-.15ex\hbox{/}\mkern-9.5mu p}

\newcommand{\Dslash}{\hbox{/\kern-.6000em D}}
\newcommand{\dslash}{\,\raise.15ex\hbox{/}\mkern-13.5mu D}

\newcommand{\raw}{\rightarrow}

\newcommand\mathC{\mkern1mu\raise2.2pt\hbox{$\scriptscriptstyle|$}
        {\mkern-7mu\rm C}}              
\newcommand{\mathR}{{\rm I\! R}}         

\newcommand\bi{\begin{itemize}}
\newcommand\ei{\end{itemize}}

\begin{document}


\begin{center}
{\large\bf Reduction, Emergence and Renormalization}  
\\
\vspace{5mm}
\end{center}

\begin{center}
Jeremy Butterfield \vspace{2mm}

{\it  Trinity College, Cambridge,  CB2 1TQ, UK: jb56@cam.ac.uk}

\end{center}

\begin{center}
Thurs 14 November 2013: \\
 in {\em The Journal of Philosophy}, volume 111 (2014), pp. 5-49; \\
based on the Nagel Memorial Lecture, Columbia University, New York: 2 April 2013
\end{center}

\begin{abstract}
\noindent In previous work,  I described several examples combining 
reduction and emergence: where reduction is understood {\em \`{a} la} Ernest  
Nagel, and emergence is understood as behaviour or properties that are novel (by some  
salient standard). Here, my aim is again to reconcile reduction and emergence, for a  
case which is apparently more problematic  than those I treated before: renormalization.

Renormalization is a vast subject. So I   
confine myself to emphasizing how the modern approach to renormalization (initiated by  
Wilson and others between 1965 and 1975), when applied to quantum field theories, 
illustrates both Nagelian reduction and emergence. My main point is  
that the modern understanding of how renormalizability is a generic feature of quantum field 
theories at  
accessible energies gives us a conceptually unified family of Nagelian reductions. 

That is  
worth saying since philosophers tend to think of scientific explanation as only explaining  
an individual event, or perhaps a single law, or at most deducing one theory as a special  
case of another. Here we see a framework in which there is a {\em space} of theories endowed  
with enough structure that it provides a family of reductions.

 \end{abstract}

\newpage

\tableofcontents

\newpage

\section{Introduction}\label{intro}

\subsection{Defending two morals}\label{twotwo}
Renormalization is undoubtedly one of the great  
topics---and great success stories---of twentieth-century physics. Also it has strongly  
influenced, in diverse ways, how physicists conceive of physical theories. So it is of  
considerable philosophical interest. I propose to honour the memory, and the philosophical  
legacy, of Ernest Nagel  by relating it to his account of inter-theoretic relations,  
especially reduction.\footnote{This paper is based in part on the Nagel Memorial Lecture,  
Columbia University, 2 April 2013. I am very grateful to Columbia University for the honour  
of giving this lecture; and for gracious hospitality and discussion. I also thank Sidney and  
Yvonne Nagel for heart-warming information about Ernest Nagel's life.}

I confess at the outset that I have an axe to grind: or more playfully, a hobbyhorse to  
ride. In previous work (especially 2011, 2011a), I argued that reduction and emergence are  
compatible. I took reduction of theories {\em \`{a} la} Nagel: as deduction of the reduced  
theory, usually using judiciously chosen definitions of its terms (bridge-laws), from the  
reducing one. And I took emergence as behaviour or properties that are novel (by some  
salient standard). In claiming this compatibility, I was much indebted to discussions by  
both philosophers and physicists that emphasized the importance of limiting relations  
between theories; especially the work in recent years by Batterman, Berry and  
Kadanoff.\footnote{For example: Batterman (2002, 2010), Berry (1994), Kadanoff (2009,  
2013). It will be clear below, starting in Section \ref{prosp}, that I interpret Nagelian 
reduction broadly, in particular as not requiring formalised languages or deductions. This 
means there is less dispute between me and some ``anti-Nagelian'' authors, such as Batterman 
himself, than there might seem.} 

Thus one of my leading ideas was that reduction and emergence are often combined by  
one theory being deduced as a limit of another, as some parameter $N \raw \infty$: the novel  
behaviour occurs at the limit. This was one of the four main Morals of my (2011), which I  
labelled `(Deduce)'. Another Moral was that the limit $N = \infty$ is not physically real, 
but an idealization, and that what  
is physically real is a logically weaker, yet still vivid, novel behaviour   that occurs  
already for finite $N$.  Thus I labelled this Moral, `(Before)', meaning `emergence (in a  
weaker form) before getting to the limit'.

I illustrated these two Morals with four main examples (and  
urged that there were many others).\footnote{I will not discuss my other two Morals: they 
were about supervenience being scientifically  
useless, and about the description of the system becoming unrealistic for very  
large $N$. In fact, I think these Morals also hold good more widely. But in Section 
\ref{Batt}, I will touch on another previous topic: whether  
the limit $N \raw \infty$ is mathematically singular, and whether this causes trouble for  
reduction.} One main example was drawn from thermal physics: phase transitions, like 
boiling, freezing and melting. If, as usual, phase transitions are taken to require 
singularities of  
thermodynamic quantities, then under very general conditions, a statistical mechanical 
description of a phase transition, i.e. a description in terms of the system's microscopic 
constituents,  requires that there be infinitely many such constituents (Kadanoff 2009 
Section 1.5, 2013 Section 2.2). Thus with $N$ as the number of constituents, the limit $N 
\raw \infty$ is called `the  
thermodynamic limit''. Our obtaining, in this limit, a description of a phase transition  
illustrates (Deduce). But the description is surely an idealization, since a boiling kettle 
contains a finite number of atoms---illustrating (Before).\footnote{Fine recent discussions 
include: Batterman (2011), Kadanoff (2009,  
2013), Menon and Callender (2013) and Norton (2012,2013).}

In this paper, I propose that  renormalization also illustrates  (Deduce) and (Before).  But 
I do not aim just to pile up illustrations. Renormalization is of special  
interest, for two reasons. The first relates to my opening remark, that it is of great 
physical importance. So we should consider how it bears on  
philosophical accounts of reduction, and related ideas like explanation.

 The other reason arises from the discussion of phase transitions. There is a class of phase 
transitions called `continuous phase transitions' (or  
`critical phemonena'), which are hard to understand quantitatively, because they involve  
many length scales. Indeed it was only by using the modern  
approach to renormalization (initiated by Wilson and others between 1965 and 1975) that they 
were quantitatively understood (even for classical, not just quantum, systems). Some authors 
have argued that these examples cause trouble for  reduction, or at least for  
Nagel's concept of it; (e.g. Batterman 2002, 2010, 2011). But I think Nagelian reduction, 
understood broadly, and thus my reconciling Morals, apply here also (my 2011, Section 7; 
Bouatta and Butterfield 2011).

However, I will not rehearse that dispute again. I propose instead to discuss  
renormalization as applied to quantum field theories about sub-atomic  
physics, rather than to  theories in statistical mechanics, about solids, liquids and gases. 
This is a vast subject: even bigger than  renormalization's description of phase transitions 
in statistical  mechanics. And I will limit myself to showing how the modern  
approach to renormalization, when  
applied to quantum field theories, illustrates Nagelian reduction. My
main claim will be that the modern understanding of how renormalizability is a generic  
feature of quantum field theories, at the sorts of energy we can access in particle  
accelerators, gives us a conceptually unified family of Nagelian reductions. (This claim  
will be developed in Sections \ref{twoapp}, 4.2.1 and \ref{manna1}.)   So here, the emphasis  
will be on illustrating the Moral (Deduce), as much as (Before).

This endeavour has three motivations, apart from the pleasure of honouring Nagel's legacy.  
First, philosophers tend to think of scientific explanation as only explaining an individual  
event, or perhaps a single law, or at most deducing one theory as a special case of another.  
Here we see a framework in which there is a {\em space} of theories endowed with enough  
structure (called the {\em renormalization group}) that it provides a family of reductions.  
Besides, each reduction will illustrate the Morals, (Deduce) and (Before). In fact, the  
parameter $N$ that goes to infinity will be distance (or equivalently: the reciprocal of  
energy); and the
emergent behaviour that is deduced in the infinite-distance, i.e. zero-energy, limit will be  
the renormalizability of the theory.

The second and third motivations arise from the current philosophical discussion about the  
modern approach to renormalization. As I mentioned, this approach applies equally to 
statistical  
mechanical theories of solids, liquids and gases, and to quantum field theories about 
sub-atomic  
physics. But philosophical discussion has emphasized the former (especially, reduction vs.  
emergence in continuous phase transitions): suggesting that it is now worth scrutinizing the  
latter. Agreed: there is also reason to resist this change of focus. The quantum field  
theories of interest here, viz. interacting  theories in four spacetime dimensions, are {\em  
not} mathematically well-defined: at least not yet! So one might say that philosophical  
scrutiny should wait until such time as they are well-defined.  But I will later (starting  
in Section \ref{heurist}) urge that despite these theories' lack of rigour, the time is ripe  
for philosophically  assessing them.

On the other hand, some of the philosophical literature about renormalization (including 
that about interacting quantum  
field theories) is ``anti-Nagelian''. For some authors hold that the modern  
approach to renormalization, and-or associated ideas like universality and effective field  
theories, make trouble for Nagelian reduction, and-or for broader doctrines like  
reductionism. I have already mentioned Batterman (2002, 2010, 2011). Also, Bain (2013, 
2013a) holds that effective  
field theories make trouble for Nagelian reduction; and Cao and Schweber (1993) hold that 
the modern approach to  
renormalization and effective field theories prompt pluralism about ontology and  
anti-reductionism in methodology.  So this is my third motivation: I will want to reply,  
albeit briefly, to these authors.

\subsection{Prospectus: Nagel endorsed}\label{prosp}
To help frame what follows,  I will first state the plan of the paper, and then briefly  
adumbrate a Nagelian view of reduction.

 In Section \ref{rnnintro}, I announce my main  
claim: that the dwindling contribution, at lower energies, of non-renormalizable  
interactions amounts to a set of Nagelian reductions. Then I urge that we can and should  
philosophically assess quantum field theories that are not mathematically well-defined. In  
Section \ref{tradmodNagel}, I sketch the traditional and modern approaches to  
renormalization. With that exposition in hand, Section \ref{Nag} returns to philosophy, and  
argues for the main claim just mentioned: that renormalizability at the energies we can  
access fits well with Nagel. Finally, in Section \ref{2ancill} I respond to some of the  
anti-Nagelian discussions mentioned at the end of Section \ref{twotwo}.\\

Turning to Nagel's account of reduction: I admire, even endorse, this account, as modified  
by Schaffner (1967, 1976, 2006). But I will not try to review, let alone defend, the  
details; for two reasons. First, Nagel's  stock-price is rising. Recent defences of Nagel's  
account, and similar accounts such as Schaffner's, include e.g. Endicott (1998), Marras  
(2002), Needham (2009), Dizadji-Bahmani, Frigg and Hartmann (2010). And  
Schaffner has recently given a masterly review of this literature, and defence of his own  
account (2013). In short, Nagel hardly needs {\em my} endorsement. Second, I have elsewhere  
given details of my endorsement (2011a, Sections 2 and 3). Some of those details will  
surface in Sections \ref{Nag} and \ref{2ancill}. But for the moment, I just set the stage by  
recalling the bare bones of the Nagelian account I endorse. I will begin, in (i) to (iii) 
below, by describing reduction as a logico-linguistic relation between theories. But then I 
will emphasize, in (1) and (2), that there is {\em no} requirement that the theories' 
languages. or the deductions of the reduced theory's claims, or the definitions of its 
terms, must be formalized.

Recall the initial idea. Theories are taken as sets of sentences closed under deducibility;  
so the intuitive idea of reduction---that a theory $T_1$ reduces another $T_2$, if $T_2$ is  
a part of $T_1$---becomes the idea that $T_2$ is a sub-theory of $T_1$. That is: as a set of  
sentences, it is a subset of $T_1$. But Nagel modifies this in three main ways. It will be  
clear that modifications (i) and (ii) answer accusations that deducibility is too strong a  
notion to express reduction of one theory to another; and that (iii) answers the accusation  
that it is too weak.

\indent (i):  $T_2$ may well have predicates, or other vocabulary, that do not occur in  
$T_1$. So to secure its being a sub-theory, we need to augment $T_1$ with sentences  
introducing such vocabulary, such that $T_2$ is deducible from $T_1$ as augmented. For a  
predicate, the simplest, and so most usually discussed, form for such a sentence is a 
definition  
in logicians' sense, i.e. a universally quantified biconditional, specifying the predicate's  
extension by an open sentence in the vocabulary of $T_1$. In short: we allow that $T_2$ is  
deducible, not from $T_1$ on its own, but from $T_1$ augmented with a set of judiciously  
chosen sentences (`bridge-laws'): Nagel (1961, pp. 354-358; 1979, pp. 366-368).

\indent (ii): What is deducible from $T_1$ may not be exactly $T_2$, but instead some part,  
or close analogue, of it. There need only be a strong analogy between $T_2$ and what  
strictly follows from $T_1$. Nagel called this {\em approximative reduction} (1979, pp.  
361-363, 371-373).

\indent (iii): These modifications, (i) and (ii), still allow there to be aspects of $T_2$,  
even aspects that are essential to its functioning as a scientific theory, that are not  
captured by any corresponding aspects of $T_1$. For example: they allow that a bridge-law  
could be so long, say a million pages, so as to defy human comprehension---and thus prevent  
$T_1$ giving us any understanding of $T_2$, or of its subject-matter. So Nagel suggested  
(1961, pp. 358-363) that in each bridge-law, the part from $T_1$ (e.g. the open sentence in  
the language  of $T_1$ that is the {\em definiens} in logicians' sense of the predicate from  
$T_2$) should play a role in $T_1$. So it cannot be a million pages long; and it cannot be a  
very 
heterogeneous disjunction.

I endorse this Nagelian account. There are three main points to make, by way of clarifying  
and defending it; (setting aside details, e.g. about accommodating functional definitions,  
which I give elsewhere: 2011a, Sections 2 and 3).

(1): The first and most important point is that, although the account takes theories as sets  
of sentences, and (i)-(iii) have a logic-chopping appearance, there is no requirement at all  
that the language of the theories, or the notion of deducibility, be formalized (let alone  
first-order). Nagel and his supporters (including Schaffner et al. listed above) of course  
know that scientific theories are not formalized, and are hardly likely to be, even in  
mathematized subjects like physics. But that does not prevent one undertaking to assess  
whether there are relations of reduction between theories (in Nagel's sense; or indeed, in  
another one). The informality merely makes one's specification of the theories in question  
somewhat vague (as are (ii) and (iii) themselves, of course); and so one's ensuing  
assessments are correspondingly tentative.

Thus in particular: there is no requirement that (i)'s `definitions' that enable a deduction  
to go through must (a) be only of predicates; or (b) use as compounding operations only the  
Boolean and quantificational ones familiar from elementary logic. They can perfectly well  
use advanced mathematical operations: e.g. limiting operations, as emphasized in Section  
\ref{twotwo}'s discussion of (Deduce). Nor is there any requirement that (ii)'s deductions 
be formalized: they need only be valid by the (informal!) standards of mathematics and 
physics.

We  can already see how this Nagelian account can fit with the Morals, (Deduce) and  
(Before). The idea is: $T_2$ with its emergent behaviour corresponds to an $N = \infty$  
limit of $T_1$. In some cases, it is not exactly true: nor even, true by the lights of  
$T_1$. But (Before) holds: a weaker yet still vivid version of the novel behaviour occurs at  
finite $N$, and strictly follows from $T_1$.\footnote{Two clarifications. (1): I agree that 
the tenor of Nagel's writings---and of his times---also suggests a narrower, and I admit 
implausible, account requiring that bridge laws be biconditionals, and even that theories, 
and so deductions, be formalized. (Thanks to Bob Batterman and Jim Weatherall for 
emphasizing this.) (2): Agreed, the Nagelian account uses the  
syntactic view of theories, which is usually opposed to the semantic view. But I contend  
that  the syntactic view, with no requirement of formalization, can describe perfectly well  
the phenomena in scientific theorizing that advocates of the semantic view tout as the  
merits of models. Indeed, I think this is hardly contentious. For to present a theory as a  
set of models, you have to use language, usually (at least in part) by saying what is true  
in the models; and on almost any conception of model, a model makes true any logical  
consequence of whatever it makes true. Thus I happily join Nagel in accepting the syntactic  
view of theories---while of course admitting there remains room for debate: for the most 
recent exchange, cf. Glymour (2013) and Halvorson (2013). But cf. (3) below, for a critique 
of the syntactic view which is more  
radical than the semantic view.}

(2): Second: By saying above `Nagelian account', and citing only Nagel himself, I do not  
mean to tie my colours exclusively to his mast. In fact, I am inclined to concur with  
Schaffner's revisions, giving what Schaffner (1977) first dubbed the GRR (`general reduction  
replacement') model of reduction. For example, Schaffner's GRR model develops (ii)'s notion  
of analogy, as a way of expressing approximate reduction. For details, I recommend Schaffner  
(2013): it situates the GRR model in the landscape of the reception of Nagel's proposals  
(Section II) and discusses  partial reductions in detail, with an optics example (Sections  
IV to VI). But my claims in this paper (especially Sections \ref{Nag} and \ref{2ancill})  
will not turn on differences between Nagel himself and Schaffner.

(3): Points (1) and (2) have been irenic: they present the Nagelian account as a broad  
church. But on another controversy, I dig my heels in. Thus some reject the idea of a  
scientific theory (whether on the syntactic, or the semantic, view) as a useful category for  
the philosophical analysis of science. Besides, some support this rejection by taking as  
their examples the mathematically ill-defined quantum field theories that will be our topic  
from Section \ref{rnnintro} onwards; saying that they should not count as theories. As an  
aspiring Nagelian, I of course disagree! But I postpone giving my reasons till Section  
\ref{thy}.

\section{Introducing quantum field theory}\label{rnnintro}
As I announced in Section \ref{twotwo}, my overall aim is to argue   
that the modern approach to renormalization illustrates Nagelian reduction; and that my  
morals (Deduce) and (Before)  hold good.  To expound this, it will be clearest first to 
emphasize the  conceptual aspects of the  
physics, without regard to Nagel (this Section and the next); and thereafter, to turn to  
philosophy, urging that the physics illustrates Nagelian reduction.

More specifically: in the rest of this Section, I will: say a bit more about my main claim  
(Section \ref{twoapp}); admit that the theories with which we are concerned are not  
mathematically rigorously defined (Section \ref{heurist}); but then hold that they are  
defined well enough that we can and should assess them philosophically (Section  
\ref{simpler}). Then in Section \ref{tradmodNagel}, I will give more details about the two  
approaches to renormalization, and how the modern approach shows that renormalizability is a  
generic feature of theories at low energies. Thereafter, I emphasize philosophical issues.

\subsection{Renormalizability at accessible energies: grist to Nagel's mill}\label{twoapp}
Renormalization is essentially a framework of ideas and techniques for taming the infinities  
that beset quantum field theories. It is usual, and it will be helpful here, to distinguish  
two approaches to the subject: a traditional one, and a modern one.

The traditional approach had its first major successes in 1947-1950, in connection with  
quantum electrodynamics (QED). In those years, figures like Dyson, Feynman Schwinger and  
Tomonaga showed that their formulations of QED, using the ideas and techniques of  
renormalization, agreed with the phenomenally accurate experiments (largely at Columbia  
University) measuring shifts in the energy-levels and the magnetic moment of an electron in  
an atom, due to vacuum fluctuations in the electromagnetic field.  After these triumphs of  
quantum electrodynamics, this approach continued to prevail for two decades. For us, the  
main point is that it treats renormalizability as a necessary condition for being an  
acceptable quantum field theory. So according to this approach, it is a piece of  good  
fortune that high energy physicists can formulate renormalizable quantum field theories that  
are empirically successful. Indeed, great good fortune: for since about 1970, high energy  
physicists have elaborated the {\em standard model}. This combines quantum electrodynamics  
with renormalizable  quantum field theories for forces other than electromagnetism---the  
weak and strong forces, that are crucial for the physics of the atomic nucleus; and for  
forty years, the standard model has had stunning empirical success.

But between 1965 and 1975, another approach to renormalization was established by the work  
of Wilson, Kadanoff, Fisher etc. (taking inspiration from ideas in the theory of condensed  
matter, i.e. liquids and solids, as much as in  quantum field theory). For us, the main  
point is that this approach explains why the phenomena we see, at the energies we can access  
in our particle accelerators, are described by a renormalizable quantum field theory.

In short, the explanation is: {\em whatever non-renormalizable interactions may occur at yet  
higher energies, their contributions to (the probabilities for) physical processes decline  
with decreasing energy, and do so rapidly enough that they are negligible at the energies  
which are accessible to us}.  Thus the modern approach explains why our best fundamental  
theories (in particular, the standard model) have a feature, viz. renormalizability, which  
the traditional approach treated as a selection principle for theories.

From a philosophical perspective, this point is worth emphasizing, quite apart from its  
scientific importance. For philosophers tend to think of scientific explanation as only  
explaining an individual event, or perhaps a single law, or at most deducing one theory as a  
special case of, or a good approximation of, another. This last is of course the core idea  
of Nagel's account of inter-theoretic reduction.

But the modern approach to renormalization is more ambitious: it explains, indeed deduces, a  
striking feature (viz. renormalizability) of a whole class of theories. It does this by  
making precise mathematical sense of the ideas of a {\em space of theories}; and a flow on  
the space, called the {\em renormalization group} (RG). It is by analyzing this RG flow that  
one deduces that what seemed ``manna from heaven'' (that some renormalizable theories are so  
empirically successful) is to be expected: the good fortune we have had is  
generic.\footnote{More generally, the idea of a flow on a space of theories, passing between  
energy regimes, has been has immensely fertile throughout physics: so fruitful that although  
Nobel Prizes tend to be awarded for experimental work, several have been awarded for  
theoretical advances associated with this idea.}

Besides, this point is, happily, {\em not} a problem for Nagel's account of inter-theoretic  
relations. For my main claim will be that it provides a conceptually unified {\em family of  
Nagelian reductions}. That is: the explanation, using renormalization group ideas, of why  
contributions to physical predictions from non-renormalizable interactions dwindle at lower  
energies,  amounts to a set of Nagelian reductions. For a renormalization scheme that  
defines a flow to lower energies amounts to a set of bridge-laws that enable a deduction  
{\em \`{a} la } Nagel, from a theory describing high-energy physics, of a low-energy theory.  
And because the same scheme shows how many similar high-energy theories  flow to  
correspondingly similar low-energy theories, we have, not just a set, but a conceptually  
unified family, of Nagelian reductions.

Returning to the Morals, (Deduce) and (Before), introduced in Section \ref{twotwo}: these  
will again be illustrated.  For from a high-energy theory with one or more 
non-renormalizable  
interactions, we deduce, by adjoining suitable bridge-laws, the low-energy theory. We take  
the parameter $N$ to be the reciprocal of energy; (or equivalently, as we shall see in  
Section \ref{tradnut}: to be distance). In the limit of zero energy (or infinite distance),  
a non-renormalizable interaction makes zero contribution. Thus renormalizability is the  
emergent behaviour which is deduced; and of course, what is physically real is the regime of  
low energies, or large but finite  distances, in which the contributions of  
non-renormalizable interactions are negligible but not exactly zero. 

I will spell out this main claim in Section \ref{manna1}, after expounding renormalization.  
And I will support it with ancillary claims, about the status of the theories concerned  
(Section \ref{thy}), multiple realizability (Section \ref{manna2}) and the idea of effective  
field theories (Section \ref{2ancill}).

But before describing renormalization (Section \ref{tradmodNagel}), I should address the  
fact that  the quantum field theories we will be concerned with---like quantum  
electrodynamics, and the other theories of the weak and strong forces that make up the  
standard model---are {\em not} mathematically rigorously defined; so that  some philosophers  
are wary of analyzing them, and some even argue that they should not count as physical  
theories. I will maintain that they are well enough defined as physical theories that a  
philosophical assessment of them is appropriate; and more specifically, that in such an  
assessment, a reasonably precise notion of physical theory---indeed, Nagel's  
notion---applies to them. I will urge this, in two stages. First: I will  urge in Sections  
\ref{heurist} and \ref{simpler}, i.e. before describing renormalization, that although they  
are ill-defined, the time is ripe for their philosophical assessment. Second: after I  
describe renormalization---and so, after we get a better sense of why these theories are  
ill-defined---I will suggest that they nevertheless pass muster as theories in Nagel's sense  
(Section \ref{thy}).

\subsection{Heuristic but successful theories}\label{heurist}
In both physics and its philosophy, there is of course a balance between heuristic, informal  
work and mathematically rigorous work. The quantum field theory community of course does  
both kinds of work; but philosophers of quantum field theory have tended to concentrate on  
the second. In this and the next Subsection, I will urge that the time is ripe for  
philosophical assessment of the first kind of work---which this paper aims to instantiate. I  
will urge this, first, in general terms; and then in Section \ref{simpler}, in relation to  
simplifications at higher energies revealed by the modern approach to renormalization.

Quantum electrodynamics, and the other theories that make up the standard model, are not  
known to be rigorously/mathematically definable. In short: they are all {\em interacting  
quantum field theories}, the interaction being, for example, between an electron and the  
electromagnetic field (due to the electron's electric charge). But after more than seventy  
years of effort, we still do not know how to rigorously define their central theoretical  
notion, viz. a path integral that gives amplitudes (i.e. complex square roots of  
probabilities) for various quantum processes---except in some special cases. (The special  
cases involve certain types of interaction, and-or a reduced number of dimensions of space,  
and-or imagining there is in fact no interaction, e.g. that all particles have zero electric  
charge, so that the electromagnetic interaction is ``turned off''---called a {\em free}, as  
against an {\em interacting}, theory.)\footnote{\label{rigorref}{For a glimpse of these  
issues, cf. e.g. Jaffe (1999, 2008), Wightman (1999).}}

What to do? Since the 1930s, the main strategy adopted has been to calculate each amplitude  
using a perturbative expansion. This is a power series, i.e an infinite sum  $\sum g^n A_n$  
with terms including successive powers of a number $g$. The broad idea is that:\\
\indent \indent (i): $g$ is usually small, in particular less than one (so that $g^n \raw 0$ 
as $n  
\raw \infty$), which helps successive terms to get smaller, and the series to converge; and  
\\
\indent \indent  (ii): physically, $g$ is a {\em coupling constant}, whose value encodes the 
strength  
of interaction between two fields: e.g. the electric charge of an electron, encoding the  
strength of interaction between the electron field and the electromagnetic field.

 But there are several severe troubles, conceptual as well as technical, about these  
perturbative expansions. The main one is that even if $g$ is small, the other factor in the  
$n$th term, viz. $A_n$, is typically {\em infinite}. This is quantum field theory's 
notorious  
problem of  infinities: which, as we will see, is addressed by renormalization. Why the  
$A_n$ are infinite, and how renormalization addresses this by introducing a cut-off and then  
analysing what happens when the cut-off tends to a limit, will be taken up in Section  
\ref{tradnut} et seq. For the moment, I just confess at the outset that the overall problem  
of infinities will {\em not} be fully solved by renormalization, even by the modern approach  
(Section \ref{modnut}). The infinities will be tamed, even domesticated: but not completely  
eliminated.

And yet, it works! For example: in quantum electrodynamics even the first few terms of the  
perturbative series give predictions that match experimental results (like those in the  
famous Columbia experiments of Foley, Kusch, Lamb, Retherford et al.) upto eleven  
significant figures; (cf. Feynman 1985, pp. 6-7, 115-119; Schweber 1994, p. 206f.; Lautrup  
and Zinkernagel 1999; Bricmont and Sokal (2004, p. 245)). This is like predicting the  
measured diameter of the USA to within the width of a human hair.

So the situation combines mathematical {\em lacunae}, even embarrassments, with supreme  
empirical success: an odd situation. In any case, for more than seventy years, the quantum  
field theory community has divided its labours between two kinds of work: (i) developing the  
heuristic formalisms of perturbation series etc.
(of course often using rigorous mathematics to do so); and (ii) pursuing a mathematically  
rigorous definition of the central notion, the path integral, and allied notions. Of course,  
the distinction between (i) and (ii) is vague. There is really a spectrum of kinds of work,  
and plenty of synergy between the different kinds; although, physics being a practical and  
empirical subject, far more quantum field theorists do work at the first heuristic end of  
the spectrum than at the second rigorous end.

In recent decades, philosophers of science who
have approached quantum field theory have tended to emphasize the second kind of work:  
especially a framework called `algebraic quantum field theory'. That is understandable:  
their temperament and training of course emphasizes rigorous argument, and one would not  
want philosophical discussions to be turn on proposals from physics that are fallible and-or  
temporary, e.g. because they depend on some specific model.

But we philosophers certainly need to engage with the first kind of work. It is not just  
that it rules the roost in physics. Also, it harbours a wealth of results  and methods which  
we can be almost certain are now permanently established; and yet which are conceptually   
subtle, and so cry out for philosophical assessment---even before ``all the theorems are  
in''. And  the material summarized in Section \ref{tradmodNagel}---the main ideas of  
renormalization---is among that wealth.

It may at first seem  rash to say, after Kuhn and his ilk, that we can be almost certain  
that some features of quantum field theory are now permanently established. But I mean it. I  
will not try to defend a general realism or cumulativism about physical theory (let alone  
science in general).\footnote{\label{sokal}{For example, cf. 
 Psillos (1999, 2009). But note that some authors suggest a role, in defending scientific  
realism, for the modern approach to renormalization. The idea is that since the  
renormalization group protects low-energy physical theories from high-energy effects, we  
can, and indeed should, believe that these theories will not be overturned by later  
discoveries about high energies; e.g. Bricmont and Sokal (2004, p. 252-254). I shall return  
to this in Section \ref{CSB}.}}

But I do want to cite a specific argument, developed by Steven Weinberg in a series of  
articles, and summarized in the opening Chapters of his magisterial treatise (1995, pp.  
xx-xxi, 1-2, 31-38; and Chapters 2 to 5; 1999, pp. 242-247): which is hardly known among  
philosophers of science, even philosophers of quantum field theory.\footnote{So far as I  
know, the only detailed discussion is Bain (1999): who construes the argument as an example  
of demonstrative induction.} Weinberg shows that any theory combining the principles of  
special relativity and quantum mechanics, and with a plausible locality property (viz. the  
cluster decomposition property), must at low energies take the form of a quantum field  
theory. Weinberg's motivation for developing the argument is not just novel pedagogy, i.e.  
to teach students an approach to quantum field theory other than the usual one of quantizing  
classical fields. The argument also shows that the framework of quantum field theory would  
still stand, even if one day we have to throw out our specific currently favoured theories,  
based on quantizing classical fields (including quantum electrodynamics). As he says:  `If  
it turned out that some physical system could not be described by a quantum field theory, it  
would be a sensation. If it turned out that the system did not obey the rules of quantum  
mechanics and
relativity, it would be a cataclysm' (p. 1).

Agreed: any such argument, however brilliant, is only as persuasive as its premises; and the  
Kuhnian, or more generally the sceptic, may deny that the principles of special relativity  
and quantum mechanics are now permanently established. Rebutting that denial would of course  
require the general arguments I have just
ducked out of attempting. But I cannot resist urging, against the Kuhnian and sceptic, that  
a theory achieving accuracies like one part in
$10^{11}$ is getting so much right about nature that much of what is claimed by its central  
principles will be retained.

\subsection{Simplifications at higher energies}\label{simpler}
There is also another reason to be optimistic about diving in to philosophically assess  
heuristic quantum field theory. In brief, the point is that some interacting quantum field  
theories are in much better mathematical shape than others. Indeed, some are in good enough  
shape that it is reasonable to hope they {\em will} be rigorously defined; so that a  
philosophical assessment, even now, is not foolhardy. More precisely: the modern approach to  
renormalization (Section \ref{modnut}) classifies (the perturbative expansions of) quantum  
field theories in part by how they behave at successively higher energies, and it turns out  
that some of our current theories are very well-behaved at high energies. The main case is  
the theory of the strong force (`quantum chromodynamics': QCD). Its
good behaviour is that as the energy gets higher, the strong force gets weaker (according to  
the perturbative analyses we are now capable of). That is, the theory tends towards a theory  
in which the ``particles'' (more precisely: excitations of a quantum field) do not ``feel  
each other'' (more precisely: evolve in time independently of each other). Such a theory is  
called {\em free}, meaning `non-interacting'. So this limiting behaviour for successively  
higher energies is called {\em asymptotic freedom} (Wilczek 2005). This striking  
simplification of the theory makes it reasonable to hope that it will be defined rigorously,  
i.e. independently of perturbative analyses.

I should emphasize that on the other hand, quantum electrodynamics gets more badly behaved  
as the energy gets higher. That is unfortunate: and not just for our physical understanding,  
but also because it fosters a misleading impression among philosophers. For there are three  
obvious factors that make philosophers tend to think of quantum electrodynamics, when the  
topic of quantum field theory is mentioned.

\indent (i): It was historically the first quantum field theory to have its infinities  
tamed, in the sense of being shown renormalizable, by the traditional approach to  
renormalization (Section \ref{tradnut}). (But as mentioned in Section \ref{heurist}, here  
`tamed' does not mean `wholly eliminated'; and nor are the infinities eliminated by the  
modern approach. This is of course a central aspect of the theories being so far  
ill-defined.) \\
\indent (ii): The electromagnetic force is familiar from everyday life, unlike the weak and  
strong forces.\\
\indent (iii): {\em Prima facie}, quantum electrodynamics is a much simpler theory than our
quantum theories of the weak and strong forces. That is: all these theories' central  
postulate is a Hamiltonian. (This is the function encoding the various fields' contributions  
to energy, that dictates how they evolve in time. It is roughly equivalent to instead  
specify a related function, the `Lagrangian'.) The Hamiltonian of quantum electrodynamics is  
much simpler than that of these other two theories. Besides, following (ii) above: its form  
is familiar from classical electromagnetic theory, i.e. Maxwell's equations. Thus I say  
`{\em prima facie}' because in studying these theories, it is the Hamiltonian that one first  
meets and analyzes. It takes  considerably more study---historically, it was a stupendous  
achievement---to show that at high energies, the comparison of simplicity can be reversed:  
that the boot is on the other foot, in that quantum chromodynamics tends to a
free theory, while quantum electrodynamics' behaviour gets worse and worse.

These three factors make quantum electrodynamics the archetype, or salient example, of a  
quantum field theory, for philosophers and other non-specialists. That is understandable,  
even reasonable. But this theory's bad behaviour at high energy, and its thereby being  
(probably!) {\em not} rigorously definable, has the unfortunate consequence of giving  
philosophers the impression that probably all interacting quantum field theories are doomed  
to not being rigorously definable. That is an unfair extrapolation: it is reasonable to hope  
that some such theories, even important theories that we believe to describe nature (viz.  
quantum chromodynamics), are rigorously definable (Besides, asymptotic freedom is not the  
only kind of good behaviour at high energy that makes this hope reasonable: two others are  
called `conformal invariance' and `asymptotic safety'.)

So much by way of motivating a philosophical assessment of interacting quantum field  
theories. Enough said: be it wise or rash, what follows is an exercise in that {\em  
genre}.\footnote{A complementary discussion of the timeliness of such assessments is in  
Bouatta and Butterfield (2012, Section 2.2). For examples of the views of theoretical  
physicists, both heuristic and mathematical, about the prospects for interacting quantum  
field theories, or at least asymptotically free ones, to be rigorously definable, cf. (i)  
the exchange between  Gross and Jaffe in Cao (1999, pp. 164-165); (ii) Jaffe (2008).}

\section{Renormalization}\label{tradmodNagel}
Renormalization  is a vast subject, and I can only scratch the surface. In this Section, I  
will just sketch some ideas of both the traditional  and the modern approaches (Sections  
\ref{tradnut} and \ref{modnut}); following Baez's helpful introductions  (2006, 2009). As 
announced  
in Section \ref{twoapp}, the overall idea will be that while the traditional approach took  
renormalizability as a selection criterion for theories, the modern approach explains it as  
a generic feature at accessible energies.\footnote{\label{Aitch}{Further details are in a  
companion paper (Butterfield 2013). For slightly more technical introductions, I  recommend:  
(i) Wilson's {\em Scientific American} article (1979) and Aitchison (1985)'s introduction to  
quantum field theory, especially its vacuum, which discusses renormalization in Sections  
3.1, 3.4, 3.6, 5.3, 6.1; (ii) Teller's philosophical introduction (1989); and as surveys  
that include some of the history: (iii) Kadanoff (2009, 2013), emphasizing renormalization  
in statistical mechanics; (iv) Cao and Schweber (1993) and Hartmann (2001), emphasizing 
renormalization in quantum field theory---to which I will return in Section \ref{CSB}.}}

\subsection{The traditional approach}\label{tradnut}
\paragraph{3.1.1: The task: corrections needed}: Consider the task of calculating the  
strength of a source from the measured force felt by a test-particle. In classical physics,  
this is straightforward. Consider a classical point-particle acting as the source of a  
gravitational or electrostatic potential. There is no problem about using the measured force   
felt by a test-particle
at a given distance $r$ from the source, to calculate the mass or charge (respectively) of  
the source particle. Thus in the electrostatic case, for a test-particle of unit charge, the  
force $F$ due to a source of charge $e$ is given, in appropriate units, by $F = e/r^2$  
(directed away from or towards the source, according as the charges of the test-particle and  
the source are of the same or opposite sign). We then invert this equation to calculate that  
the source's charge is: $e  =  F.r^2$.

This straightforward calculation of the source's mass or charge---in the notation of Section  
\ref{heurist}, the coupling constant $g$---does not work in quantum field
theory! There are complicated corrections we must deal with. These depend on the energy  
and-or momentum with which the test-particle approaches the source. A bit more exactly,  
since of course the test-particle and source are equally minuscule: the corrections depend  
on the energy or momentum with which we theoretically describe, or experimentally probe, the  
system I called the `source'. We write $\mu$ for this energy (and  say `energy', not `energy  
or momentum'). These corrections to the coupling constant $g$, depending on $\mu$, will be  
centre-stage in both the traditional and the modern approaches to renormalization.

So we write $g(\mu)$ for the {\em physical coupling constant}, i.e. the coupling constant  
that we  measure: more exactly, the coupling constant that we calculate from what we {\em  
actually} measure, in the manner of $g = g(F)$ above, in the simple electrostatic example.  
Thus the notation registers that
$g(\mu)$ is a function of $\mu$.

The need for corrections then means that $g(\mu)$ is not the same as the {\em bare coupling  
constant}, $g_0$ say, that appears in the theory's fundamental equations (like $e$) in the  
electrostatic example). But we expect $g(\mu)$ to depend on $g_0$. So we write $g(\mu)  
\equiv g(\mu, g_0)$. And our task is to invert this equation, writing $g_0 = g_0(g(\mu))$,  
and so to assign a value to $g_0$ that delivers back the measured $g(\mu)$ for the various  
energies $\mu$ with which we observe the system: or more generally, for the various energies  
$\mu$ for which we are confident of the value of $g(\mu)$.

This statement of our task needs to be refined. In fact, we need three refinements, which I  
will treat in order:\\
\indent (i) introducing a cut-off;\\
\indent (ii) letting the cut-off go to a limit;\\
\indent (iii) allowing for extra terms in the Hamiltonian.\\
But if we can succeed in the task, as refined, we will say that the interaction in question,  
or the theory that describes it, is {\em renormalizable}. After these refinements, I will  
report that the theories in the standard model are indeed renormalizable.

\paragraph{3.1.2: The cut-off}:
This task is daunting, because as announced in Section \ref{heurist}, our technique for  
calculating an amplitude, viz. power series, seems to break down: the factor $A_n$ in the  
$n$th term $g^n A_n$ is typically infinite. The reason is that $A_n$ is an integral over  
arbitrarily high energies. To try and get a finite answer from our formulas, the first thing  
we do is impose a {\em cut-off}. That is: we replace the upper limit in the integral,  
$\infty$,  by a high but finite value of the energy, often written $\Lambda$: we require by  
{\em fiat} that the contribution to the integral from higher energies is zero. So with $k$ 
representing energy, we require that $\int^{\infty}_{\Lambda} dk \; ... \equiv 0$. (There 
are other less crude ways to secure a finite answer---called {\em regularizing} the 
integrals---but I will only consider cut-offs). Thus  the  physical coupling constant, 
$g(\mu)$, is a function, not only  
of the bare coupling constant $g_0$ and of $\mu$ itself of course, but also of the cut-off  
$\Lambda$:
\be\label{gmu}
g(\mu) \equiv g(\mu, g_0, \Lambda).
\ee
There are two immediate points to make.

First: in quantum theory, energy (and momentum) are like the reciprocal of distance; in the  
jargon, `an inverse distance': energy $\sim$ 1/distance. (And so distance is like an inverse  
energy.)  So high  energies correspond to short distances; and so to short wavelengths and  
to high frequencies. So the cut-off  $\Lambda$ corresponds, in terms of distance, to a  
cut-off at a small distance $d$. That is: by imposing the cut-off to get finite answers, we  
are declaring that  any fields varying on scales less than $d$ do not contribute to the  
specific process we are calculating.

Second: It turns out to be possible, and  is very convenient, to express all dimensions in  
terms of length. Thus we can also trade in the energy-scale $\mu$ for an inverse length, say  
$\mu \sim 1/L$ where $L$ is a length. So we re-express the physical coupling constant as a  
function of $L$: we will use the same letter $g$ for this function, so that we write $g(L)  
\equiv g(\mu)$. Thus eq. \ref{gmu} becomes:
\be\label{gL}
g(L) \equiv g(L, g_0, d).
\ee

Thus  our task can now be stated as follows. We are to measure $g(L)$  (better: to calculate  
it from what we really measure, like the force $F$ in the simple electrostatics example) and  
then invert eq. \ref{gL}, i.e. write $g_0 = g_0(g(L), d)$,  so as to calculate which value  
of the bare constant would give the observed $g(L)$, at the given $d$.

\paragraph{3.1.3: Letting the cut-off $d$ go to zero}:
Broadly speaking, the exact value of the cut-off is up to us.\footnote{Agreed: for some  
perturbative analyses of some problems, the physics of the problem will suggests a range of  
values of $d$ that are sensible to take. That is: the physics suggests that no phenomena on  
scales much smaller than $d$ will contribute to the process we are analysing.} But of  
course, the theory and its predictions should be independent of any human choice. And if our  
theory is to hold good at arbitrarily short lengths (arbitrarily high energies), we expect  
that $g_0$ goes to a limit, as $d$ tends to zero (at least at some appropriate $L$: such as  
the observed $L$).

If this limit exists and is finite, i.e. $\in \mathR$, we say: the interaction or theory  
with we are concerned  {\em is finite}. But most successful quantum field theories are {\em  
not} finite. The paradigm case is QED, for which the limit is infinite. That is: for  
arbitrarily high cut-offs, the bare coupling constant $g_0$ becomes arbitrarily high. 
Mathematically,  
this is like elementary calculus where some function $f(x)$ tends to infinity as $x$ tends  
to infinity,
e.g. $\lim_{x \raw \infty} \surd x = \infty$. But of course this last is `just' the infinity  
of pure mathematics. But here we face a {\em physically real infinity} viz. as the value of  
the bare coupling constant.

The consensus, on the traditional approach to renormalization, is that this physically real  
infinity is {\em acceptable}.  Accordingly, the adjective `renormalizable', with its  
honorific connotations, is used. That is: If $g_0$ tends to a limit, albeit perhaps  $\pm  
\infty$, we say the theory is {\em renormalizable}. So in particular: QED is renormalizable  
in this sense, though not finite.\footnote{But I should add that despite this consensus,  
most physicists would admit to some discomfort that the bare constant should be infinite in  
the continuum theory. Thus  
great physicists like Dirac have been very uncomfortable (cf. the citations in Cao (1997,  
pp. 203-207)); and Feynman himself calls renormalization `a dippy process' and `hocus-pocus'  
(1985, p. 128).}

\paragraph{3.1.4: Allowing for extra terms}:
It turns out that to write down a renormalizable theory, we may need to add to the  
Hamiltonian function (equivalently; Lagrangian function) one or more  terms to represent  
extra fields, or interactions between the given fields, even though we believe the bare  
coupling constant for the extra fields or interactions are zero. The reason is that the  
interaction might have a non-zero {\em physical} coupling constant at some scale $L$; i.e.  
$g(L) \neq 0$.

So now we should  generalize the notation slightly to reflect the fact that there are   
several, even many, coupling constants to consider; as well as  several, even many, possible  
interactions (terms in the Hamiltonian). So suppose that
there are in all $N$ physical coupling constants, $g_1(L), g_2(L), ... g_N(L)$, occurring in  
the various terms/interactions in our theory.

We similarly generalize slightly our  definition of renormalizability. A theory which  
secures that each bare coupling constant goes to a limit as $d$ tends to zero, by using  
either (i) no extra terms, or (ii) at most a  finite number of them, is given the honorific  
adjective: {\em renormalizable}.

The consensus, on the traditional approach, is that renormalizability in this sense is a  
necessary condition of an acceptable theory.  Clearly, this seems a reasonable view. That  
is, renormalizability seems a mild condition to impose, since its definition has  
accommodated a succession of complications about the idea of assigning a bare coupling  
constant: we have had to allow for dependence on the energy $\mu$, on the cut-off, and for  
extra terms.

\paragraph{3.1.5: Our good fortune}:
So much by way of explaining the idea of renormalizability. How do the quantum field  
theories we ``believe in'', or ``take seriously'' fare? That is: are the theories which are  
our best descriptions of the electromagnetic, weak and strong forces, renormalizable in the  
sense just discussed? Yes they are; though they are not finite.

This circumstance seems a piece of great good fortune. At least, echoing the remarks just  
above: we are likely to feel it is a relief after our: (a) having to admit that we can so  
far only define the theory perturbatively (Section \ref{heurist}); and (b) having to make   
corrections to bare coupling constants which turned out to require a succession of 
complications: (i) dependence on $\mu$ ($L$), (ii) a cut-off that then goes to infinity, and  
(iii) extra terms.

But we will now see that according to the modern approach to renormalization, this great  
good fortune is not so surprising. In a certain sense, {\em renormalizability is generic} at  
the low-ish energy scales we can access.

\subsection{The modern approach}\label{modnut}
The key initial idea of this approach is that instead of being concerned with good limiting  
behaviour as the cut-off $d \raw 0$, we instead focus on how $g(L)$ varies with $L$.

Indeed, if we envisage a number of coupling constants, say $N$ for $N$ possible  
interactions, then the ``vector'' of coupling constants $(g_1(L),..., g_N(L))$ represents a  
point in an $N$-dimensional space; and as $L$ varies, this point flows through the space.    
And accordingly: if we envisage a theory as given by a Hamiltonian which is a sum of terms  
representing different possible interactions, then this space is a space of theories.  
Jargon: we say the coupling constants {\em run}, and the flow is called the {\em  
renormalization group flow.}

This simple idea leads to a powerful framework, with rich consequences not just in quantum  
field theory, but in other branches of physics, especially statistical mechanics and the  
theory of condensed matter. But I shall concentrate on how it explains why a theory about  
phenomena at the low (or low-ish!) energy scales we can access, is renormalizable. That is:  
it explains  the good fortune reported at the end of Section \ref{tradnut} as being generic.  
This discussion will introduce some jargon, indeed ``buzz-words'', such as `fixed points'   
and `universality'.

\paragraph{3.2.1: Good fortune explained: non-renormalizable terms dwindle at longer  
distances}:
 There are of course various  controversies about explanation. But it is surely   
uncontroversial that one very satisfying way to explain the good fortune reported at the end  
of Section \ref{tradnut} would be to show: not merely that some given theory is  
renormalizable; but that {\em any} theory, or more modestly, any of a large and-or generic  
class of theories, is renormalizable.  Such an argument would demonstrate that our good  
fortune was ``to be expected''. (Admittedly, such an explanation, whether for a single  
theory, or for a class of them, will  have to make some other assumptions about the theory  
or theories: a point I will stress in Section 3.2.2 below. So it is only relative to those  
assumptions that the good fortune is explained, and to be expected.)

This is  what the modern approach to renormalization gives us, with its idea of a space  of  
theories, on which there is a flow given by varying the energy-scale $\mu$ or $L$.

More precisely and modestly: this approach does not show that any of a large and-or generic  
class of theories has, at the comparatively low energies and large length-scales we can  
access, literally {\em no} non-renormalizable terms. Rather, the approach shows that for any  
such theory---``with whatever high-energy behaviour, e.g. non-renormalizable terms, you  
like''---the non-renormalizable terms dwindle into insignificance as  energies become lower  
and length-scales larger. That is, in Section \ref{tradnut}'s notation: the physical  
coupling constant for non-renormalizable terms shrinks. For such terms: as $\mu \raw 0$  
(i.e. $L \raw \infty$), $g(\mu) \equiv g(L)$ $\raw 0$.

The explanation is based on a simply stated criterion for when a theory is renormalizable:  
more precisely, for when a term is renormalizable. It is a matter of the dimension (as a  
power of length) of the bare coupling constant in the term. Namely: this dimension needs to  
be less than or equal to zero.  The criterion is due to Dyson, and is sometimes called {\em  
Dyson's criterion}.

 More precisely: suppose  that the bare coupling constant $g_0$ has  
dimensions of length$^D$. Then it turns out that the corresponding physical coupling  
constant $g(L)$ will scale roughly like $L^{-D}$. That is:
\be\label{gLscale}
g(L)/g_0 \sim (L/d)^{-D} \; .
\ee
Thus if $D > 0$, the exponent on the right-hand side will be negative; so when $L$ is very  
small, i.e. much smaller than $d$, the right hand side is very large. That is: the physical  
coupling constant will be large compared with the bare one. That is a sign of bad  behaviour  
at small distances $L$, i.e. high energies. At least, it is bad in the sense that the large  
coupling constant will prevent our treating the interaction represented by the term as a  
small perturbation. So it is unsurprising that such a term is non-renormalizable in the  
sense that Section \ref{tradnut} sketched.

But now, instead of considering  the case of  very small $L$ (so that a non-renormalizable  
term's positive exponent $D$ makes for a large physical coupling constant): {\em look at the  
other side of the same coin.} That is: when $L$ is  much larger than the cut-off $d$, and $D  
> 0$ (i.e. the term in question is non-renormalizable), then the right hand side of eq.  
\ref{gLscale} is {\em very small}. That is: the physical coupling constant is very small. So  
at large distances, the non-renormalizable interaction is weak: ``you will not see it''.

There are two main points I should make about this explanation, before addressing the  
Nagelian themes of explanation and reduction (Section \ref{Nag}). The first point is about  
how non-trivial the {\em explanans}, i.e. eq. \ref{gLscale}, is. The second point will  
somewhat generalize the discussion, from a physical not philosophical viewpoint; and will  
introduce some jargon.

\paragraph{3.2.2 Decoupling high-energy behaviour}:
That at large distances, a non-renormalizable interaction is weak follows immediately from  
eq. \ref{gLscale}. But that does not make it obvious! A good deal of theory needs to  
be assumed in order to deduce eq. \ref{gLscale}. After all, there is of course no {\em a  
priori} guarantee that interactions that are strong at short distances should be weak at  
long distances. To show this ``decoupling'' of high-energy behaviour from the low-energy  
behaviour was a major achievement of Wilson and many other physicists, e.g. Symanzik (1973),  
Applequist and Carazzone (1975). I will not go into details, but just  remark that it can be  
shown under very general conditions, even within the confines of a perturbative analysis.

\paragraph{3.2.3 The renormalization group flow}:
So far, my talk of the renormalization group flow has been restricted in two ways, which I  
need to overcome.

\indent \indent  (a): A flow can have a {\em fixed point}, i.e. a point that is not moved by  
the flow: think of sources and sinks in elementary discussions of fluid flow. For the  
renormalization group flow, this would mean a set of physical coupling constants  
$(g_1(L),..., g_N(L))$ that is unchanged as the length-scale $L$ increases further. Jargon:  
the behaviour of the system is {\em scale-invariant}: ``you see the same  
behaviour/theory/physical coupling constants, at many different length-scales''. This can  
indeed happen: the kind of phase transition mentioned in Section \ref{twotwo}, i.e. 
continuous phase transitions, provides vivid examples of this. Such a point is called an 
{\em infra-red fixed point}. Here, `infra-red' is used on analogy with light: infra-red 
light has a longer  
wavelength, lower frequency and lower energy, than visible light.

\indent  \indent  (b):  So far, we have had in mind one trajectory, maybe leading to a fixed  
point. But many trajectories might lead to the same fixed point; or at least enter and  
remain in the same small region of the space. If so, then the  `vectors' $(g_1(L),...,  
g_N(L))$ at diverse early points on a pair of such trajectories representing dissimilar  
theories lead, as $L$ increases, to the same fixed point, or at least to the same small  
region, and so to similar theories. That is: when you probe at low energies/long distances,  
``you see the same or similar physical coupling constants''. Jargon: This is called {\em  
universality}. And the set of `vectors' that, as $L$ increases, eventually lead to the given  
fixed point, is called, on analogy with elementary discussions of fluid flow, the point's  
{\em basin of attraction}.  

But note that universality should really be called `commonality'  
or `similarity', for two reasons. (i): There can be different fixed points, each with their  
own basin of attraction; and (ii) for a given basin, the vector of physical coupling  
constants does not encode {\em everything} about the system's behaviour, so that systems  
with the same vector  will not behave indistinguishably. But jargon aside: Section  
\ref{manna2} will urge that universality is essentially the familiar philosophical idea of  
multiple realizability.

Finally, I can summarize this Subsection's main point, that non-renormalizable interactions  
dwindle at large length-scales, by combining the jargon I have just introduced with the  
previous jargon that a {\em free} theory is a theory with no interactions. Namely: the  
infra-red fixed point of a theory {\em all} of whose interaction terms are nonrenormalizable 
is a  
free theory.

\section{Nagelian reflections}\label{Nag}
So much by way of introducing quantum field theories (Section \ref{rnnintro}) and   
renormalization (Section \ref{tradmodNagel}). In this Section and the next, I turn to this  
material's bearing on philosophy. As announced in Section \ref{prosp}, I will undertake:  
first, the positive task of fitting this material to Nagel (this Section); and then, the  
negative  task of replying to some anti-Nagelian discussions (Section \ref{2ancill}). Both  
Sections will connect with Section \ref{intro}'s endorsement of a broadly Nagelian account  
of reduction, and its Morals (Deduce) and (Before).

In this Section, I first urge that although our quantum field theories are not well-defined  
(Section \ref{heurist}), they pass muster as theories in Nagel's sense (Section \ref{thy}).  
Then in Section \ref{manna2}, I will argue that universality (cf. Section 3.2.3) is  
essentially the familiar philosophical idea of multiple realizability. Then in Section  
\ref{manna1}, I make my main claim: the explanation, using renormalization group ideas, of  
why contributions to physical predictions from non-renormalizable interactions dwindle at  
lower energies (cf. Section 3.2.1),  amounts to a family of Nagelian reductions.

\subsection{Endorsing the idea of a theory}\label{thy}
In recent decades, various philosophers  have for various reasons criticized or even  
rejected the notion of theory: even in relaxed versions that make no requirement of a formal  
language, and whether construed syntactically or semantically. They say that the notion of a  
scientific theory over-emphasizes the linguistic, logical or semantic aspects of science at  
the expense of other important, but less formal or tidy aspects. For example: the non-formal  
aspects of explanation and confirmation; the role of analogy and metaphor; the creation and  
application of models (in scientists', not logicians', sense!); instrumentation, experiment  
and simulation; and more generally, the embedding of a theory in wider practices, both  
cognitive (research and pedagogy) and non-cognitive, in laboratory, lecture-room and society  
at large.

Obviously, these aspects, both individually and taken together, militate against `theory'  
being precisely defined: just because they are informal and complicated. And they are so  
various that they favour no single revision of the notion of a theory. Their collective  
effect is rather to suggest `the scientific theory' is no longer a useful---or at least, not  
a crucial---category for the philosophical analysis of science.\footnote{\label{unity}{Nor  
is theory the only category, traditionally central to the philosophy of science, to be thus  
questioned. The notion of a law of nature, or even a law of a specific theory, has also been  
rejected; e.g. Cartwright (1983), Giere (1995).}}

Besides, this philosophical critique has been supported by historical and sociological  
studies: including studies of the very theories I have discussed, viz. quantum field  
theories. Here, I have in mind  the work of such authors as Galison (1997) and Kaiser  
(2005). For example:  Kaiser, in his monograph about Feynman diagrams and thus also the  
history of quantum electrodynamics, urges that `the scientific theory' is not a useful  
historiographical category (2005, pp. 377-387). As you might guess, his reason is  
essentially that these theories are not rigorously defined---prompting the suggestion that  
approximation techniques, like perturbation series, Feynman diagrams etc., are ``all there  
is to the physics''.

Obviously, I cannot here reply to this general philosophico-historical critique of theory;  
nor even to Kaiser and the other authors about quantum field theories. So I will just say  
what I consider the main justification for using `theory' as I have done repeatedly in  
Sections \ref{rnnintro} and \ref{tradmodNagel}. This reply will lead us back to the  
syntactic conception of theories, and so return us to Nagel's proposed analysis of  
reduction.

My main point is that we should distinguish  theory in general, from specific theories. That  
is: we must distinguish two issues. They are superficially similar. For both have two  
corresponding parts: for each of them  concerns:\\
\indent \indent (Def): whether `theory' can  be precisely defined, faithfully to its root  
meaning;\\
\indent \indent (Use): whether `theory'  is a useful unit or category for the philosophical  
analysis of science.\footnote{Of course (Def) and (Use) are related in various ways. In  
particular, since precision aids analysis, one expects a `Yes' to (Def) to count in favour  
of a `Yes' to (Use); but since after all, analysis can be illuminating even without all its  
terms being precisely defined, a `Yes' to (Def) is not a pre-requisite of a `Yes' to  
(Use).}\\
But within both (Def) and (Use), we must distinguish two issues: viz. whether `theory'  
means:\\
\indent \indent \indent \indent (Gen): theory in general, or \\
\indent \indent \indent \indent (Spec): a specific theory.

Obviously, the developments in general philosophy of science just adumbrated concern (Gen).  
They suggest that the  notion of a physical (or more generally: scientific) theory cannot be  
precisely defined, and-or is not a useful unit or category for analysis. And also obviously:  
Nagel, by the very act of proposing an analysis of inter-theoretic reduction, rejects these  
suggestions; (even setting aside how the rest of his account of science deploys the notion  
of theory). I join him in this, though I have ducked out of justifying the rejection.

But my concern in Sections \ref{rnnintro} and \ref{tradmodNagel}  has been with (Spec): with  
some specific  theories, viz a handful of  quantum field theories. So agreed: I need to  
address the questions (Def) and (Use) for each such theory---though I do not propose to do  
so here. Again, this leads back to the question  whether we should be instrumentalist about  
these theories.

Here, I want just to emphasize the evident but important point  that this paper's main  
enterprise does {\em not} depend on retaining the general notion of theory, i.e. on joining  
Nagel (and me) in rejecting the recent developments' `anti-theory' suggestions. That is:   
one can be sceptical or agnostic about the general issues, i.e. one can say `No' or `Maybe'  
to (Def) and-or (Use) under meaning (Gen)---while confidently saying `Yes' to (Def) and-or  
(Use) for a specific physical theory.

For consider: the terms with which one might hope to define the general notion of physical  
or scientific theory are very different from those with which one might hope to define a  
specific theory. The first group of terms would include logico-linguistic items such as  
`sentence', `model', `deductive closure', and maybe also more philosophical items such as  
`confirmation', `explanation' and `domain of application'. But the second group would  
include items specific to the theory concerned, such as  `energy', `momentum' and  
`Lagrangian'. So one might well think ill of the first group of terms, i.e. be sceptical or  
agnostic about usefully defining the general notion of theory, while also thinking well of  
the second group, i.e. while being confident that a specific physical theory can be  
precisely defined; (though of course for interacting quantum field theories, `precisely  
defined' must be taken as logically weaker, i.e. less demanding, than rigorous definition  
within pure mathematics). Indeed: ever since Section \ref{rnnintro}, I have often identified  
a quantum field theory with its Hamiltonian (or Lagrangian), i.e. with a specification of  
the quantum fields and their postulated interactions. So in this usage of `theory' for a  
specific theory, as in (Spec), one might well confidently say `Yes' to (Def) and-or  
(Use).\footnote{{\label{FDsplit}}{This claim was in play in the differing attitudes of  
Feynman and Dyson during the early 1950s to quantum electrodynamics (QED), in the light of  
its bad behaviour at high energies. Roughly speaking, Feynman is less demanding. He held  
that the results and calculational techniques of perturbative QED, including the eponymous  
diagrams, are so fruitful and accurate that one should ``run with it''. (He of course would  
not care whether we should label this body of doctrine with the honorific word `theory'.)  
Dyson, trained as a pure mathematician, was more demanding. He was disheartened by the  
apparent impossibility of a rigorous definition of the theory.  (For the history of this  
disagreement, cf. Schweber (1994, pp. 564-572) and Kaiser (2005, pp. 175-195, 246-248, 358),  
who calls it `the Feynman-Dyson split'.) Thus my claim puts me with Feynman: and I am happy  
to call the body of doctrine a `theory', and so to disagree with Kaiser's conclusion (ibid.,  
pp. 377-387) that there is no theory hereabouts.}}


\subsection{Universality is multiple realizability}\label{manna2}
In Section 3.2.3, I introduced `universality' as jargon for the idea that dissimilar  
theories might have similar infra-red behaviour: in particular, the same infra-red fixed  
point. Here I wish to: (i) point out that universality is essentially the familiar  
philosophical idea of multiple realizability, and (ii) make two ancillary comments.

As to (i), I fortunately do not need to worry about the exact definition of multiple  
realizability: i.e. about whether multiple realizability requires more than a property's  
being disjunctive, for example by the disjunction being sufficiently heterogeneous  
(according to some, maybe vague, standard), or by the property being functional (i.e.  
quantifying over properties). For the examples usually given of dissimilar theories having  
similar infra-red behaviour will undoubtedly count as examples of multiple realizability:  
the theories  are strikingly dissimilar, and the infra-red behaviour strikingly similar.

In fact, the examples usually given are from statistical mechanics, rather than quantum  
field theory. As mentioned in Section 3.2.3, they concern continuous phase transitions.  The  
dissimilar theories are of utterly different quantities in utterly different systems. For  
example, one such quantity is the difference of the densities of water and steam, in a  
system that is a mixture of water and steam; another quantity is the density difference, in  
a mixture of two phases of liquid helium; a third is the magnetization of a piece of iron or  
nickel. Despite these systems being so disparate, they show some similar infra-red  
behaviour. Besides, this behaviour is striking because it is quantitative, exact and  
concerns an arcane quantity. Namely, it concerns the values of the exponents (called  
`critical exponents') in the power laws governing these quantities' values at temperatures  
close to that at which the continuous  phase transition occurs.\footnote{More specifically:  
such a transition occurs only under specific conditions, in particular at a specific  
temperature, the {\em critical temperature} $T_c$ (which is {\em not} itself universal).  
Close to this temperature and other conditions, the value of some quantity, $v(Q)$ say, is  
given  by a power of the difference between the actual temperature $T$ and $T_c$: $v(Q) \sim  
|T - T_c|^p$,
where $p$ is the power (also known as: exponent). Here, $p$ might be positive, so that $Q$'s  
value is zero at the transition: this occurs for the density differences and magnetization I  
mentioned, for which $p$ is about $0.35$ (and is called $\beta$). But for other sets of  
(again, mutually disparate) quantities $Q$, their shared power law has a negative exponent   
$p$, so that the value of each quantity $Q$ diverges at the phase transition.}

So much by way of illustrating universality, and its being multiple realizability. Turning  
to my two ancillary comments: the first is in effect a note of caution to philosophers, 
about the success  
of Section \ref{modnut}'s renormalization group framework in explaining such universality:  
in particular, its correctly predicting the exponents in these power laws for countless such  
systems. This success is much celebrated, including in the philosophical literature about  
phase transitions; and rightly so.

But beware: this praise can give the impression that only with the renormalization group did  
physics cotton on to the broad idea of: \\
\indent \indent (a) starting with a description of a system including details about short  
distances (equivalently, as in Section 3.1.2: high energies); and then \\
\indent \indent (b) systematically dropping information from the description (often called  
`coarse-graining'), so as to get from a `microscopic' description to a `macroscopic' (long  
distance, low energy) one. \\
That is {\em not so}. Coarse-graining a microscopic description to get a macroscopic one has  
of course long been endemic in physics. The reasons are obvious---the microscopic  
description is often intractably complicated, involving a vast number of degrees of freedom;  
and the strategies are familiar---e.g. partitioning the state-space, especially by defining  
collective variables.

Besides: even for the striking examples above, viz. the universality of critical exponents, 
some approaches to calculating these exponents, that were developed long before the 
renormalization  
group, had considerable, albeit partial, success. That is of course unsurprising, since the  
idea of coarse-graining is so natural. But it is worth emphasizing, to avoid philosophers  
getting the impression that only with the renormalization group did physics pick up on the  
idea of multiple realizability.\footnote{\label{MFT}{I think this false impression is 
fostered by some philosophers' praise of the renormalization group; e.g. Batterman (2002, 
pp. 37-44), Morrison (2012, p. 156, p.160 (both paragraph 2)). Incidentally, the two main 
previous approaches to calculating critical exponents are mean field theory and Landau 
theory. For a glimpse of what these are and their predictive limitations,  
in the context of condensed matter, cf. e.g. Kadanoff (2009, Sections 2,3; 2013, Sections  
1.2.4-4, pp. 147-164) and Binney et al. (1992, pp. 22, 176-177, 183-184). For example, mean  
field theory implies that the value of $p$, in the previous footnote, is $0.5$, as against  
the actual $0.35$.}}

My second comment is a brief Nagelian one. Namely: I think that multiple realizability, and  
thus universality, does not cause problems for Nagelian reduction. Obviously, that is good  
news for me, since  I have endorsed a Nagelian account of reduction (Section \ref{prosp};  
and 2011a, Sections 3.1.1 and 4.1.1). I will not go into detail since my reasons derive from 
Sober  
(1999); and anyway, I have developed them elsewhere (2011a, Section  
4.1).\footnote{\label{Kim}{To summarize: Sober mostly targets Putnam's (1975) and Fodor's  
(1974) claims that multiple realizability prevents reduction, especially in the philosophy  
of mind: claims which Kim endorsed and developed into a rival account of reduction (1999, p.  
134; 2005, pp. 99-100; 2006, p. 552). For critiques of Kim, cf. e.g. Marras (2002, pp.  
235-237, 240-247) and Needham (2009, pp. 104-108). Within philosophy of physics, perhaps the  
most extended claims that multiple realizability causes trouble for Nagelian reduction are 
by Batterman, especially in his earlier work. But I postpone discussing him until Section  
\ref{Batt}.}}

\subsection{Renormalizability deduced at low energies as a family of Nagelian  
reductions}\label{manna1}
Taking the last two Subsections' pro-Nagelian conclusions in my stride, I turn to my main  
claim, announced in Section \ref{twoapp}: that the modern understanding of how  
renormalizability becomes generic (`emerges') as we consider theories at lower and lower  
energies amounts to a conceptually unified family of Nagelian approximative reductions.

The idea is clear from the details of Section 3.2.1. A renormalization scheme that defines a  
flow to lower energies (equivalently: long distances) amounts to a set of bridge-laws that  
enable a deduction {\em \`{a} la } Nagel, from a theory describing high energy (short  
distance) physics, of a low energy (long distance) theory. The fact that contributions to  
physical predictions from non-renormalizable interactions dwindle as we consider lower and  
lower
energies, means that only renormalizable terms are significant at low energies---and become  
more significant as the energy gets lower. So we have Nagelian approximative reduction. And  
my Morals (Deduce) and (Before) of Section \ref{twotwo} are both illustrated, with the  
emergent behaviour being renormalizability, and the parameter $N$ being the distance $L$ (or  
reciprocal of the energy $\mu$). Furthermore, because the same renormalization scheme shows  
how many high energy theories  flow to corresponding low energy theories,
we have, not just a set, but a conceptually unified family, of Nagelian reductions.

To sum up, I claim:
\begin{quote}
The deduction from a given theory $T_1$ that describes (perhaps using non-renormalizable  
terms) high-energy physics, of an approximately renormalizable theory $T_2$ describing low  
energy physics, 
is a Nagelian reduction. Besides: for different pairs of theories $T_1$ and $T_2$, varying  
across the space of quantum field theories, the reductive relation is  similar, thanks to a  
shared definition of the renormalization group flow, i.e. of the renormalization scheme.
\end{quote}
I will fill this out with five short remarks, rehearsing previous material. As regards the  
philosophical assessment of Nagelian reduction, the most important of these remarks are (3),  
about approximate reduction, and (4), about unity among a family of reductions.

(1): I have specified a theory by a Hamiltonian or Lagrangian. Recall the vector of physical  
coupling constants $g_1(\mu),..., g_N(\mu)$ which I first introduced in Section 3.1.4, and  
took as a point in a space of theories at the start of Section \ref{modnut}. Recall also my  
defence of this notion of a specific theory (as against the notion of a theory in general)  
in Section \ref{thy}.

(2): A renormalization scheme that defines a flow towards lower energies (a scheme  for  
coarse-graining so as to eliminate higher-energy degrees of freedom) amounts to the set of  
definitions or bridge laws ((i) of Section \ref{prosp}) needed to make the deduction of  
$T_2$ from $T_1$ go through.

(3): Since at low energies any non-renormalizable terms in $T_2$ still make non-zero, albeit  
dwindling, contributions to the theory's predictions (probabilities for various processes),  
we have here an approximative reduction ((ii) of Section \ref{prosp}); though the  
approximation gets better and better as the energy decreases.

(4): A given renormalization scheme (definition of the flow) works to show that many  
theories $T_1$ lead to approximately renormalizable low-energy theories $T_2$. This unity  
is striking. Hence this Section's title's use of the word `family': since family' connotes  
resemblance, which `set' does not.

(5): Agreed: no single renormalization scheme works to prove that all possible theories have  
dwindling non-renormalizable contributions in the infra-red. And as I have admitted: the  
proofs concerned are often not mathematically rigorous. But the various renormalization  
schemes that have been devised do not contradict one another, and in fact mesh in various  
(often complicated) ways. So it is fair to say that a generic quantum field theory has  
dwindling non-renormalizable contributions in the infra-red.

So to sum up: the modern approach to renormalization gives a stunning case of explaining  
something that is, on the traditional approach, a coincidence. The coincidence is  that the  
theory in question, e.g. quantum electrodynamics, has a feature, viz. renormalizability,  
that seems crucial to it ``making sense''. This feature turns out to hold for the whole  
standard model of particle physics, which combines quantum electrodynamics with quantum  
theories of the weak and strong forces. Thus the coincidence is so large and important that  
it can seem like manna from heaven; or more prosaically, it can seem that renormalizability  
is in some way an {\em a priori} selection principle for quantum field theories. But  
adopting the modern approach, we can deduce that what seemed manna from heaven is in a sense  
to be expected.

A final Moral. As I mentioned in Section \ref{twotwo}, I think philosophers should take 
note,  
not just of this
specific achievement, but of the general idea of a {\em space of theories}. This fosters a  
novel and more ambitious kind of explanatory project than the familiar ones of explaining an  
individual event,  or  a single law, or a theory as a part of, or a good approximation to,  
another. Namely: to explain a feature of a whole class of theories in a unified way in terms  
of the structure of the space of theories.\footnote{Nor is the renormalization group the  
only example of this idea in physics. Another example is catastrophe theory.  Roughly  
speaking: this takes a theory to be given by a potential function (similarly to it being  
given here by a Hamiltonian or Lagrangian); the space of potential functions is then endowed  
with structure such as a topology, in terms of which features of potentials such as their  
being structurally  stable are then  explained.}

\section{Ongoing controversies: effective field theories}\label{2ancill}
My claims in Sections \ref{manna2} and \ref{manna1} have a controversial edge to them. For  
some have argued that the modern approach to renormalization prompts (i) anti-Nagelian  
morals, and more broadly (ii) anti-reductionist morals. So in this final Section, I will  
reply briefly to some of these claims. I begin with Batterman, who is perhaps the best known  
of the authors concerned (Section \ref{Batt}). Then I introduce the idea of {\em effective  
field theories}: an idea which the modern approach to renormalization has fostered, and on  
which the anti-Nagelian and anti-reductionist morals of Bain, and of Cao \& Schweber, are  
based. So I first expound the idea (Section \ref{efts}) and then reply to these authors  
(Section \ref{CSB}).

\subsection{Batterman: singular limits?}\label{Batt}
Batterman has long argued that Nagel's account of reduction (more generally: inter-theoretic  
relations) does not fit explanations using the renormalization group  (cf. especially his  
2002, 2010, 2011). But I will be very brief about his views, for two (admittedly partial)  
reasons. First, he focusses on the  explanation of critical exponents in continuous phase  
transitions (cf. Section \ref{manna2}), and thus on the renormalization group in statistical  
mechanics, rather than in quantum field theories for sub-atomic physics. Second, I have  
already elsewhere given a Nagelian reply (2011, Section 7; 2011a Sections 3 and 4.1; Bouatta  
and Butterfield 2011).

Agreed: as so often in academic controversy, peace has not yet broken out! So here is a bit  
more detail, to fill out Section \ref{manna2}'s closing comment that multiple realizability  
is no trouble for Nagel. As mentioned in footnote \ref{Kim}, Kim believes it is, and  
develops his own account of reduction, which he calls a `functional model'. It takes  
reduction to include (a) functional definitions of the higher-level properties $P$ etc. and  
(b) a lower-level description of the (variety-specific) realizers of $P$ etc., and of how  
they fulfill the functional roles spelt out in (a). Batterman agrees with Kim that multiple  
realizability spells trouble for Nagel, and that Kim's model is not Nagelian. But he  
rejects Kim's model because it does not sufficiently recognize features common across 
different varieties of realizer, and so does not help explain the relative autonomy of the 
``higher level''. He even suggests programmatically that the understanding of universality 
provided by the renormalisation group (and so {\em contra} my footnote \ref{MFT}'s praise of 
mean field theory and Landau theory) could help us recognize such common features and 
explain the autonomy of the higher level---even for properties far removed from physics, 
such as pain. (For details of these views, cf. Batterman (2002, pp. 65, 67, 70-72, 73-75 
respectively).)

Of course, there is a lot more to Batterman's views; and besides, they have changed over  
time. Both these points are shown by his (2010). Its example of multiple realizability is  
Gibbs' discussion of how in his two main frameworks for statistical mechanics (called `the  
canonical ensemble' and `the micro-canonical ensemble'), different quantities are the  
analogues, as the number $N$ of constituents in the system goes to infinity (the  
thermodynamic limit), of the thermodynamic quantities, entropy and temperature. Batterman  
expounds this example in relation both to Gibbs himself, and to the connection between the  
renormalization 
group  and probability theory.\footnote{To see the idea of this connection, recall that the  
central limit theorem says, roughly speaking, that the sample averages obtained by  
independent sampling tend to a normal distribution as the sample size $N$ increases; and  
this is so for independent sampling of any of a wide class of probability distributions.  
This suggests that increasing $N$ defines a flow in the space of distributions under which  
any distribution in the class flows to the normal distribution. So in the jargon of Section  
3.2.3: the class is  a basin of attraction, and the normal distribution is a fixed point.  
Cf. also Batterman (2013, pp. 275-278).}

As regards Nagelian reduction, Batterman uses this example to make two main points. The  
first is a peace-pipe for Nagel; the second, about singular limits, less so. First, he  
agrees that where the limit---$N \raw \infty$ in our notation---is not singular, 
the Nagelian  `conception of derivational reduction will likely hold ... because we can take 
the limiting  relations as providing us with something like the bridge laws appropriate for 
Nagel-like  reduction' (2010, p. 166); and he adds that expressing these relations will 
typically require mathematical operations, not the universally quantified biconditionals of 
philosophical discussions. This peace-pipe for Nagel is offered again, with the details of 
the Gibbsian example of multiple realizability, later on (p. 174, paragraphs 3 and 4).

I of course endorse this quotation, since I think the Nagelian can and should admit multiple  
realizability and the need to go beyond biconditionals: recall that for me and my fellow  
Nagelians, there is no requirement for a formal language (cf. (1) of Section \ref{prosp}).

Second, Batterman emphasizes that the limit is often mathematically singular;\footnote{As he  
had in previous work; and as do his kindred spirits, Berry and Kadanoff, cited in Section  
\ref{twotwo}.} and in such cases, he contends, `it is best to give up talk of `reduction'  
altogether and to speak instead of `intertheoretic relations'. In this paper, Batterman's  
example of such a singular limit is his favourite one: continuous phase transitions, also  
known as critical phenomena. (I think his best exposition of this example is in his (2011,  
Sections 3 and 5).) 

The topic of singular limits is a large one, which I cannot take up; but I will make two  
comments.\\
\indent \indent (1): Batterman softens this apparently anti-Nagelian point by saying that  
the singularities concerned `are not genuine obstacles to some kind of general limiting  
(reductive?) relation between the theories after all'; and adding that in this regard, he  
has changed his view (p. 176, and footnote 9 respectively). \\
\indent \indent (2): Whether a limit is singular can be a subtle matter. It is not just that  
there are various mathematical types of singularity. Also, a physical phenomenon or set of  
phenomena can be modelled both by a formalism with a singular limit, and by one with a  
continuous limit. And  in some such cases---cases that are striking and so rightly  
emphasized by Batterman and others---the second formalism with a continuous limit is at  
least as adequate and rigorous as the first. In short: philosophers should beware of loose 
talk  
that limits are singular.\footnote{Cases where rigorous algebraic quantum theory gives a 
{\em  
continuous} limit include: (i) spontaneous symmetry breaking, both in the $N \raw \infty$ 
limit of quantum statistical mechanics, and in the $\hbar \raw 0$ limit of wave mechanics 
(Landsman 2013, Section  3); and (ii)  the emergence as $N \raw \infty$ of  
superselection (Landsman 2007, Section 6.1-6.7, my 2011, Section 6)).}

\subsection{Effective field theories}\label{efts}
I turn to explaining how the modern approach to renormalization has fostered the idea of  
{\em effective field theories}. The reason goes back to  Section 3.2.1's explanation of  
non-renormalizable contributions dwindling at long distances. I will first emphasize how  
that explanation does not depend on spacetime being a continuum (Section \ref{notcontm}); 
and then  
describe how this suggests (though it does not imply!) a `tower' of merely effective  
theories (Section \ref{only?}).

\subsubsection{Spacetime might not be a continuum}\label{notcontm}
Recall Section  3.2.1's explanation of non-renormalizable contributions dwindling at long  
distances, using the scaling equation eq. \ref{gLscale}: which, to repeat it, was
\be\label{gLscalerepeat}
g(L)/g_0 \sim (L/d)^{-D} \; .
\ee
It is clear that this explanation does not depend on our theory (with all its terms,  
including non-renormalizable ones) being true, or even approximately true, at arbitrarily  
short distances. Our theory only needs to be approximately true at suitable intermediate  
distances. More precisely: it only needs to secure eq. \ref{gLscalerepeat} holding for any  
non-renormalizable interaction at a range of scales which is wide enough to include $L$  
being sufficiently larger than the cut-off $d$, so that with the given positive dimension  
$D$ of the bare coupling constant, the left hand side of eq. \ref{gLscalerepeat} is small  
enough that we will not see the interaction. (That is, as I put it in Section 3.2.1.: the  
right hand side of eq. \ref{gLscalerepeat} is  small enough to make the left hand side small  
enough that ``you will not see it''.)

We can put the same point in more physical terms, and in terms of energies. Maybe at very  
high energies, spacetime does not behave like a continuum. But provided the theory is ``true  
enough'' at some high, maybe even inaccessible, energies in the sense that it validates eq.  
\ref{gLscalerepeat}, then we can deduce that at much lower, in particular accessible,  
energies, ``we see only renormalizable interactions''. That is: our theory's predictions  
have significant contributions only from renormalizable interactions.

In short: we can be agnostic about whether---indeed, we can even deny that---our theory  
describes physics in a continuous spacetime. All we need is that it is approximately true at  
suitable intermediate distances, in the sense just specified.

Here we meet a widespread jargon. A theory that is taken to be approximately true in a given  
regime (of energy and-or length, and-or some other parameters) is called {\em effective}.  
The adjective is  used especially when the theory is known or believed to be {\em only}  
approximately true, because it is derived from a theory with better epistemic warrant, by  
adopting  certain approximating and-or idealizing assumptions (assumptions which go beyond  
merely specifying the regime, i.e. range of parameters, concerned).

So we can sum up Section 3.2.1's explanation of what I called `our good fortune' by saying:  
from studying the renormalization group flow, we deduce (subject to the substantive   
assumptions gestured at in Section 3.2.2!)  that effective low-energy theories are  
renormalizable. This leads in to the next point.

\subsubsection{Effective theories only?}\label{only?}
Section 3.2.1's explanation prompts the rhetorical question: why worry about  
non-renormalizable terms? Although they induce bad behaviour, i.e. a large coupling, at  
short distances, this bad behavour is invisible at the larger distances we can access. So  
why not countenance non-renormalizable terms, at least for inaccessibly high  
energies?\footnote{Note the contrast with the idea, on the traditional approach to  
renormalization, that renormalizability is a necessary condition for a theory to be  
acceptable; cf. Section 3.1.4.}

Of course, the words `worry' and `countenance' are vague. What you are inclined to worry  
about, and correspondingly what you are willing to countenance, will depend on your  
background attitudes to quantum field theories: for example, on how confident you are about  
using them at high energies, and about accepting results obtained from a heuristic  
formalism, rather than by rigorous mathematical proofs. So there are bound to be several  
possible positions. Here I will develop one position, often called the {\em effective field  
theory programme} (or: approach). It is based, not on confidence about the two topics above,  
but on an opportunistic or instrumentalist attitude to being {\em unconfident} about  
them.\footnote{Recalling Section \ref{simpler}, we can already glimpse why several 
positions are possible. For there I cited results showing a realistic quantum field theory's  
(viz. QCD's) good behaviour at arbitrarily short distances, and thus better prospects for  
being rigorously definable.}

There are of course two main factors that prompt a cautious or sceptical attitude towards  
the framework  of quantum field theory.\\
\indent \indent (1): One is just that interacting quantum field theories (in four spacetime  
dimensions) are at present mathematically ill-defined (Section \ref{heurist}). \\
\indent \indent (2): The other factor is the expectation that at sufficiently high energies,  
the framework breaks down, to be replaced by a theory or theories using a different  
framework. This break-down might occur only at the vast energies
associated with quantum gravity: the replacement theory being perhaps a version of string  
theory, or some other current contender for a theory of quantum gravity. Or the break-down  
might occur at intermediate energies, energies far higher than we can (and probably: ever  
will) access, but well below those of quantum gravity: there are proposals for new  
frameworks at these energies, such as non-commutative geometry.

Either or both of these factors prompt one to be cautious about drawing from quantum field  
theory conclusions about ontology. Or rather: conclusions about the ontology of phenomena at  
very high energies, or very short distances. But these factors should {\em not} suspend all  
discussion of ontology in the light of physics, or even in the light of quantum field  
theory; for four reasons.\\
\indent \indent (a): Whatever the phenomena at very high energies turn out to be, whatever  
the theoretical framework for describing them, and whatever ontology that framework  
suggests, we have every reason to expect that the facts at those energies determine, i.e.  
subvene, the facts at the lower energies we can access.\\
\indent \indent (b):  And given the great success of quantum field theory, we have every  
reason to expect that the facts at those very high energies imply a quantum field theoretic  
description at the lower, accessible, energies.\\
\indent \indent (c): This last point can be strengthened. Recall from Section \ref{heurist}  
Weinberg's argument that physics at accessible energies must be described by a quantum field  
theory, {\em even if} the framework breaks down higher up, e.g. because of gravity. In  
short: any quantum theory that at low enough energies obeys special relativity and satisfies  
cluster decomposition (which is plausible, since it has the flavour of a locality  
assumption), must be a quantum field theory.   \\
\indent \indent (d):  Besides, whoever said that ontology concerns only ``the supervenience  
basis'', i.e. the putative set or level of facts that determine (subvene) all other facts?  
That is: there is plenty of scope for ontological discussion of supervening (``higher  
level'') facts and theories: in particular, there is scope for ontological discussion of  
quantum field theory.

But these two factors also suggest that even below the energy scale at which the entire  
framework of quantum field theory breaks down, there may, for all we know, {\em not} be any  
single quantum field theory which is more fundamental than the others, in the sense that  
each of them is derived from it by assuming extra conditions that specify the derived  
theory's regime (of energies and types of interaction considered etc.). That is: as the  
energy scale gets higher and higher (while remaining below the scale at which the entire  
framework of quantum field theory breaks down), physics could, for all we know, be described  
by a succession of quantum field theories, each of which accurately describes the phenomena  
at a certain range of energies, but becomes inaccurate above that range. And when it becomes  
inaccurate, it may also become even more badly behaved, mathematically.

This scenario is often called the {\em tower of effective field theories}. But the phrase  
can be misleading, for two complementary reasons.\\
\indent \indent (i): First, as I mentioned when defining `effective', at the end of Section  
\ref{notcontm}: the adjective is often used when the theory is known or believed to be only  
approximately true, because it is derived from another theory with better warrant, by  
adopting  approximating and-or idealizing assumptions. But note: in this scenario, the  
theories in the envisaged tower are {\em not} required to be derivable from some other  
theory: in particular, one with better warrant for being taken as exactly true  
(`fundamental'), because it also covers higher energies. Rather, each theory  is simply  
accurate in its energy range, and inaccurate beyond it.\\
\indent \indent (ii): Second: the word `tower' suggests an infinite tower. But as I noted in  
(2) above, there are good reasons (concerning  quantum gravity, if nothing else) to think  
that at {\em some} energy, the entire framework of quantum field theory breaks down. This  
implies that, considered as a programme or approach for quantum field theory, the effective  
field theory programme can, and should, take it that the tower is probably finite.

But setting aside misleading connotations: the main point here is that this scenario gets  
some support from Section 3.2.1's explanation of ``our good fortune'', viz. that any  
non-renormalizable interactions (terms), though they would be important at higher energies,  
will make a dwindling contribution to all processes, as the energy scale is reduced. For  
this explanation implies that we cannot get evidence about which non-renormalizable  
interactions, if any, operate at inaccessibly high energies. Whatever they are---and  
whatever bad short-distance behaviour they suffer (eq. \ref{gLscalerepeat}, with $L << d, D  
> 0$ so that both sides of the equation are large)---we will not see them. So why worry  
about non-renormalizable interactions (terms)? Thus for all we know, or could ever know, the  
scenario of the tower holds good: there is no fundamental quantum field theory, and various  
non-renormalizable interactions operate at various inaccessibly high energies.

So much by way of sketching the effective field theory programme. We can sum it up as urging  
that, regardless of how and why quantum field theory might break down at very high energies:  
we have no reason in theory, nor experimental data, to deny the scenario of the tower---a  
succession of theories, each accurately describing physics in its energy range, and  
inaccurate beyond it.

\subsection{Bain, Cao and Schweber: against reductionism}\label{CSB}
What should we make of the effective field theory programme, from a philosophical viewpoint?  
Of course, I must postpone a detailed discussion to another occasion. But I shall make two  
obvious, yet important, points. They also set the scene for brief replies to discussions by  
Bain, Cao and Schweber which urge, albeit in different ways, that effective field theories  
give evidence against ``reductionism''.

(1) First: the vision of the tower of effective theories bears on the broad, and maybe  
perennial, debate between instrumentalist and realist views of physical theories, and of  
science in general. For the vision is that for all we know, there is a sequence of quantum  
field theories, each accurate within, but not beyond, its own energy-range, that are {\em  
not} each derivable from some single theory (viz. as an approximation or limit describing  
that range). This `for all we know' is reminiscent of the instrumentalist's leading idea of  
the under-determination of theory by data. All the more so, when we note that in the present  
state of knowledge, no one---not even an expert in effective field theories---can show that  
there is at most {\em one} such sequence: there could be several towers, even with some of  
them sharing some  `floors' in common.

(2) Second: whatever we could eventually come to believe (whether on present evidence and  
much reflection, or on the basis of much more evidence, even evidence about inaccessibly  
high energies that we surely cannot get)  about there being such a tower of theories, and  
whatever the upshot of the broad debate between instrumentalist and realist views---my main  
claim (Sections \ref{twoapp}, 3.2.1, and \ref{manna1}) {\em is unaffected}. For we must  
distinguish `reductionism' in its various possible senses, from the claim that Nagelian  
reduction is illustrated in some, or even many, cases. Agreed: some more precise versions of  
the idea of the tower of theories would, if true, be a counterexample---and a striking and  
important one---to various precise senses of `reductionism'. But I have of course not  
defended any such sense; but only that the `flow to the infra-red' illustrates Nagelian  
reduction (in a broad sense: Section \ref{prosp}).

Indeed, {\em au contraire}: it seems that in general, the tower of theories is good news,  
rather than bad, for my main claim. For it promises more illustrations of Nagelian reduction  
than my previous defence (Section \ref{manna1}) described. For at each `floor' of the tower,  
there is an effective field theory accurately describing an energy-range, so that one can  
envisage the description it gives of the lower end of its energy-range being reduced {\em  
\`{a} la} Nagel by a renormalization scheme (flow to the infra-red) being applied to its  
description of the upper end of its energy-range.\\


With these two points in hand, I turn to reply  to: (i) Bain (2013, 2013a) and (ii) Cao and 
Schweber (1993). My reply to Bain will be very brief, since the focus is sharp: we plainly 
disagree  
about my main claim. Cao and Schweber have larger and different concerns: they discuss  
reductionism, and several other `isms', but do not mention Nagelian reduction. I will engage  
a little with these `isms', so that my reply to them will be longer. But for both (i) and  
(ii), my main emphasis will lie in (2) above: that my main claim is unaffected by the rise 
of  
the effective field theory programme. \\

\paragraph{5.3.1 Bain}  Bain (2013) reviews effective field theories, first technically 
(Sections 2 to 4) and  
then in relation to reduction and emergence (Sections 5 and 6). Similarly, in his (2013a) he 
first reviews effective field theories technically and interpretively (Sections 2 and 3), 
and then discusses emergence. The two papers are similar, and I will concentrate on (2013). 
Even so, there are many details  
I cannot address.\footnote{One pertinent example is that Bain (2013 Section 6.3) asks 
whether   
effective field theories illustrate Batterman's proposal (cf. Section \ref{Batt}) that  
reduction fails when the relevant mathematical limit is singular. On the other hand, since  
Bain's Sections 5 and 6 also discuss Cao and Schweber and their critics, my Section 5.3.2 
below will  implicitly engage with some of what he says. In any case: my thanks to Bain for  
correspondence; indeed, I believe we now agree.} I will only register my disagreement with 
Bain's statements:
\begin{quote}
[Because] the steps involved in the construction of an effective field theory typically  
involve approximations and heuristic reasoning ... [and] identifying the high-level  
variables ... it will be difficult, if not impossible, to reformulate the steps involved in  
the construction of an effective field theory ... in the form of a derivation (2013, Section  
6.1 end, p. 246-247).
\end{quote}
and 
\begin{quote}
The dynamical laws of an effective field theory and its high-energy theory are different,  
and a difference in dynamical laws entails a difference in theorems derived from those laws.  
Thus an effective field theory is not a sub-theory of its high-energy theory; hence one  
cannot say that an effective field theory reduces to its high-energy theory, on this  
[Nagelian] view of reduction (2013, Section 6.2(a), p. 247).
\end{quote}
(There are similar statements in his (2013a): e.g. p. 262 (a) (b) and note 7, p. 264 
paragraph 3, p. 266 paragraph 2.) 

The reasons for my disagreement are obvious from Section \ref{prosp}'s account of broad  
Nagelian reduction. As to the first statement: (1) of Section \ref{prosp} emphasized that  
there is no requirement of formalization. And I submit that according to the standards of  
informal mathematical or physical reasoning, many a specification of a renormalization  
scheme, and implementation of it so as to define a flow in the space of theories, would  
count as a derivation. In particular, its reliance on approximations and heuristics, and  
even on insight or creativity to identify the correct variables to manipulate, does not  
prevent its being a derivation.\footnote{At the end of Section 5.3.2 below, this last point, 
that  
Nagelian reduction of course allows that formulating a bridge law can require insight or  
creativity, will recur.}
 
As to the second statement: (ii) of Section \ref{prosp} emphasized Nagel's idea of  
approximative reduction (which is also emphasized in Schaffner's GRR account). That is: what  
is deducible from $T_1$ (here, the high-energy theory) may not be exactly $T_2$ (here, the  
effective theory), but only some part, or close analogue, of it. The standard example is  
that you cannot deduce Galileo's law of free fall (that acceleration during free fall is  
constant) from Newton's theory, which says that the acceleration increases slightly as the  
body gets closer to the centre of the Earth and so feels a stronger gravitational force.  
Thus, since Newton's theory surely does reduce Galileo's law, Nagel says that reduction only  
requires deduction within error bars; or in more positive words, deduction of a corrected  
version of $T_2$. Similarly, say I, for effective field theories: i.e. for the  
dwindling contribution of non-renormalizable terms at long distances/low energies. \\

\paragraph{5.3.2 Cao and Schweber}   As I mentioned in footnote \ref{Aitch}: Cao and 
Schweber (1993) is a survey of  
renormalization in quantum field theory. Much of it surveys the history; (it draws on Cao  
(1993) and Schweber (1993), in a contemporaneous anthology, Brown (1993), emphasizing  
history). But this historical survey has a philosophical sting in the tail.  Cao and  
Schweber begin their thirty-page discussion of `philosophical ramifications' (1993, Section  
4, pp. 69-90) by saying that the modern approach to renormalization, and effective field  
theories, imply
\begin{quote}
a pluralism in theoretical ontology, an anti-foundationalism in epistemology, and an  
anti-reductionism in methodology ... in sharp contrast with the neo-Platonism implicit in  
the traditional pursuit of quantum field theorists, which ... assumed that, through rational  
(mainly mathematical) human activities, one could arrive at an ultimate stable theory of  
everything (p. 69).
\end{quote}
They then develop the three `isms' listed, in subsequent subsections (e.g. Section 4.1 on  
pluralism). The discussion is wide-ranging: for example, it touches on the theory of meaning  
for scientific terms (p. 75) and the philosophy of pure mathematics (p. 81-83). It is also  
vigorous, indeed broad-brush. For example: (i) about their own position:
\begin{quote}
The empiricist position in epistemology that is supported by the recent developments in  
renormalization theory is characterized by its anti-essentialism and its  
anti-foundationalism, its rejection of a fixed underlying natural ontology expressed by  
mathematical entities, and its denial of universal, purely mathematical truths in the  
physical world (p. 77).
\end{quote} 
and (ii) against the opposition:
\begin{quote}
The latest example of such an overly grandiose and totalizing conception of physical theory  
is the search for the theory of everything by superstring theorists (p. 74).
\end{quote}

Fighting talk! Admittedly, as I said in (2): some precise versions of the tower of theories  
would, if true, be a counterexample to various precise senses of `reductionism'. All parties  
can agree to that. The task must be to state such precise versions and senses, and to assess  
whether the technical scientific developments instantiate them. As I said, I myself must  
postpone that to another occasion. Here it suffices to make three points. The first is  
critical of Cao and Schweber. The second is a {\em caveat}, and leads into the third, which  
returns to my main claim.

First: Unsurprisingly, Cao and Schweber's controversial claims have attracted criticism. I  
shall consider three main replies: Huggett and Weingard (1995, 172 and 187-189), Hartmann  
(2001, Section 4.2, 297-299) and Castellani (2002, 263-265).\footnote{\label{eft}{Other  
philosophical discussions of effective field theories include Weinberg (1999, pp. 246-250),  
Redhead (1999, pp. 38-40) and Cao (1997, Section 11.4 pp. 339-352).}}
Among Cao and Schweber's three `isms', they focus on pluralism and anti-reductionism. As I 
read them,\footnote{I say `as I read them' since I am summarizing disparate  
discussions. For example, Huggett and Weingard distinguish different strengths of the EFT  
approach (of course placing Cao and Schweber among the stronger versions). Hartmann  
addresses the three isms {\em seriatim} in the light of his fine preceding surveys of 
renormalization in general (his Section 2.2-2.3) and of the interplay between theories, 
models and effective field theories, in particular in QCD (his Section 3).} their main  
reply is that in the present state of knowledge, we have no compelling reason, even for  
energies for which we can be confident of the quantum field theory framework (and so:  
independently of considerations of quantum gravity), to believe in what in (1), at the start 
of this Section, I called the  
`vision': viz. a tower of theories that are {\em not} each derivable from some single 
theory, as  
an approximation describing physics at its own energy range. Thus Castellani sums up:
\begin{quote}
The fact that current quantum field theories are now seen as effective field theories does  
not imply any specific thesis about the existence of a final theory ... The effective field  
theory approach does not imply that the idea of a theory being more fundamental than another  
is meaningless (2002, p. 264-265).
\end{quote}
I of course agree.

Second, a {\em caveat}: Several passages in Cao and Schweber appear to admit this point,  
i.e. what I just called these critics' `main reply'. Thus it is not really clear what Cao  
and Schweber believe the development of physics (up to twenty years ago, i.e. 1993) {\em  
implies} or at least {\em supports} (their words; p. 69), by way of `philosophical  
ramifications'---as against what it merely suggests. Some examples:\\
 \indent (i): They concede that derivation of our successful theories (i.e. the standard  
model) from some single (indeed: renormalizable and unified) theory describing higher  
energies is possible; (p. 65, start of (i)). Similarly, in some of their other  
contemporaneous work. Thus Schweber  writes:
\begin{quote}
[One can take] the viewpoint that quantum field theory will ultimately yield a fundamental  
theory. If a complete renormalizable theory at infinitely short distances  were available,  
one would be able to work one's way up [i.e. to longer distances, lower energies: JNB] to  
the effective theory at any larger distance in a totally systematic way by integrating out  
the heavy fields of the theory (1993, p. 154).
\end{quote}
\indent \indent (ii) They concede that their opponents, i.e. believers in such a single or  
fundamental  theory,  have tenable replies to some of their arguments. For example: their  
question why renormalization as a topic makes for such fruitful exchanges between quantum  
field theory and statistical physics (pp. 72-73) can be answered in a `realist-essentialist'  
way by appealing to the `unity of physical phenomena on the ontological plane, and/or the  
universality of physical truths on the epistemological plane' ((p. 73).

Third: In one of their more precise statements  (under `pluralism in ontology', their 
Section  
4.1), Cao and Schweber take as distinctive of their own view (and so as rejected by `monist'  
or `reductionist' opponents) the claim that it is impossible
 \begin{quote}
to infer the complexity and novelty that emerge at the lower energy scales from the  
simplicity at higher energy scales, {\em without any empirical input} [their emphasis]. The  
necessity, as required by ... EFT, of an empirical input into the theoretical ontologies  
applicable at lower energy scales is fostering [a picture in which the energy scales are]  
layered into quasi-autonomous domains, each layer having its own ontology and associated  
``fundamental'' laws (p. 72).
\end{quote}
But I think most of us, even many self-styled monists and reductionists, would happily agree 
(a)  
that empirical input is needed; and (b), concerning the second half of the quotation, that 
the  
lower energy scales, or more generally special sciences (`layers'), are largely autonomous,  
with their own distinctive concepts, laws and methods.\footnote{Thus this quotation provides  
another passage exemplifying my {\em caveat} just above: it is not really clear to me what  
Cao and Schweber intend.}

The point here is of course wider (and thankfully, more elementary!) than discussions of  
quantum field theory. One sees it in basic discussions of Nagelian bridge laws. To formulate  
the bridge laws that enable an inference of the reduced theory (`the complexity and  
novelty') from the reducing (`the simplicity'), one of course needs empirical  
input---namely, to suggest what in the former might be associated with what in the latter.  
And that need remains, even if: (a) after the formulation and then successful reduction, the  
bridge law is glossed as a definition in logicians' sense (e.g.  for a predicate: a  
universally quantified biconditional); and-or (b) over time, terms' meanings change so that  
the bridge law comes to seem analytic or a matter of convention. 

Besides, as many authors stress: one often needs, not just empirical input, but high  
scientific creativity. Examples are legion. A standard example is Maxwell's formulation of a  
bridge law identifying electromagnetic waves with visible light. He was led to it by  
deducing from his electromagnetic theory the speed of the waves, and noticing that it  
matched the measured speed of light. Without that empirical input, even the genius of  
Maxwell might not have formulated the bridge law.  Another standard example is formulating  
which statistical mechanical quantities to associate with thermodynamic quantities (as  
discussed by Nagel himself, and Batterman cf. Section \ref{Batt}): again, empirical input  
and scientific creativity are required in order to bring the two theories into consistent  
contact, and {\em a fortiori}, to deduce the one from the other. Yet another standard  
example is elementary chemistry. Without empirical input and creativity, even the genius of  
Schr\"{o}dinger (or indeed, the entire team of founding fathers of quantum mechanics!) could  
surely not have inferred from the many-body Schr\"{o}dinger equation (with, say, the nucleus  
treated as fixed, and including spin terms) the existence of the elements, let alone some of   
their chemical behaviour, such as the  structure of the periodic table.

This need for empirical input and for creativity is, I take it, uncontroversial.  And  
applied to our topic of effective field theories, it is well made by Castellani in her  
summing up:
\begin{quote}
The effective field theory approach does not imply anti-reductionism ...the effective field  
theory schema, by allowing definite connections between theory levels, actually provides an  
argument {\em against} [anti-reductionism]. A reconstruction (the way up) [i.e. derivation  
of lower energy behaviour: JNB] is not excluded, even though it may have to be only in  
principle (2002, p. 265).\footnote{This last sentence  echoes Castellani's previous summary  
(p. 263): `the effective field theory approach provides a level structure of theories, where  
the way a theory emerges from another [a notion she has earlier glossed, largely in terms of  
Nagelian reduction: JNB] is in principle describable by using RG methods'. Again, this  
returns us to my main claim.}
\end{quote}
In any case: controversial or not, this point returns us back to my main claim, that a  
family of renormalization group trajectories flowing to the infra-red (`the way up' in  
Castellani's and Schweber's jargon) gives a unified family of reductions, in my broad  
Nagelian sense; (Sections \ref{twoapp}, 3.2.1, and \ref{manna1}). As I emphasized in (2) at  
the start of this Subsection, this claim is not tied to any form of reductionism; and in  
particular, not to denying the tower of effective theories. It says `Here, we find unified  
families of reductions'; but nothing like `Here is a single grand reduction of all of  
particle physics'. \\ 



\newpage

{\em Acknowledgements}:--- I am very grateful to Jonathan Bain, Bob Batterman, Tian Yu Cao, Elena Castellani, Sebastien Rivat and Jim Weatherall for generous comments on a previous 
draft; and to audiences at Columbia University, New York, Munich and Cambridge. While I have incorporated many of these comments, I regret that, to keep my main argument clear and 
short, I could not act on them all. I am also grateful to Nazim Bouatta for teaching me about renormalisation and related topics.This work was supported by a grant from the Templeton World Charity Foundation.  
The opinions expressed in this publication are those of the author and do not necessarily  
reflect the views of Templeton World Charity Foundation. \\ \\

\section{References}

Aitchison, I. (1985), `Nothing's plenty: the vacuum in quantum field theory', {\em  
Contemporary Physics} {\bf 26}, p. 333-391.

Applequist, T. and Carazzone, J. (1975), `Infra-red singularities and massive fields', {\em  
Physical  Review} {\bf D11}, pp. 2856-2861.

Baez, J. (2006), `Renormalizability': (14 Nov 2006):\\
 http://math.ucr.edu/home/baez/renormalizability.html

Baez, J. (2009), `Renormalization made easy': (5 Dec 2009): \\  
http://math.ucr.edu/home/baez/renormalization.html

Bain, J. (1999), `Weinberg on quantum field theory: demonstrative induction and  
underdetermination', {\em Synthese} {\bf 117}, pp. 1-30.

Bain, J. (2013), `Effective Field Theories', in Batterman, R. (ed.) {\em The
Oxford Handbook of Philosophy of Physics}, Oxford University Press, pp.
224-254.

Bain, J. (2013a), `Emergence in Effective Field Theories', {\em European Journal
for Philosophy of Science}, {\bf 3}, pp. 257-273. (DOI 10.1007/s13194-013-0067-0)

Batterman, R. (2002), {\em The Devil in the Details}, Oxford University Press.

Batterman, R. (2010), `Reduction and Renormalization',  in A. Huttemann and G. Ernst, eds.  
{\em Time, Chance, and Reduction: Philosophical Aspects of Statistical Mechanics}, Cambridge  
University Press, pp. 159--179.

Batterman, R. (2011), `Emergence, singularities, and symmetry breaking', {\em 
Foundations of Physics}, {\bf 41}, pp 1031-1050.

Batterman, R. (2013), `The tyranny of scales', in R. Batterman, ed., {\em The Oxford  
Handbook of Philosophy of Physics}, Oxford University Press, pp. 255-286.

Bedau, M. and Humphreys, P. eds. (2008), {\em Emergence: Contemporary Readings in 
Philosophy and Science}, MIT Press.

Berry, M. (1994), `Asymptotics, sngularities and the reduction of theories', in  D. Prawitz,  
B. Skyrms and D. Westerdahl (eds), {\em Logic, Methodology and Philosophy of Science IX},  
Elsevier Science: pp. 597-607.

Binney, J., Dowrick, N, Fisher, A. and Newman, M. (1992), {\em The Theory of Critical  
Phenomena: an introduction to the renormalization group}, Oxford University Press.

Bouatta, N. and Butterfield, J. (2011), `Emergence and Reduction Combined in Phase  
Transitions', in J. Kouneiher, C. Barbachoux and D.Vey (eds.), {\em Proceedings  of  
Frontiers of Fundamental Physics 11} (American Institute of Physics). Available at: http://  
philsci-archive.pitt.edu/8554/ and at: http://arxiv.org/abs/1104.1371

Bouatta, N. and Butterfield, J. (2012), `On emergence in gauge theories at the 't Hooft  
limit', forthcoming in {\em The European Journal for Philosophy of Science}; http://philsci-archive.pitt.edu/9288/

Bricmont, J. and Sokal, A. (2004), `Defence of modest scientific realism', in {\em Knowledge  
and the World: Challenges Beyond the Science Wars}, ed. M. Carrier et al. (Springer), pp.  
17-45; reprinted in Sokal's collection (2008), {\em Beyond the Hoax: science, philosophy and  
culture}, Oxford University Press, pp. 229-258; page references to reprint.

Brown, L. ed. (1993), {\em Renormalization: from Lorentz to Landau (and Beyond)}, Springer.

Brown, L. and Cao. T. (1991), `Spontaneous breakdown of symmetry: its rediscovery and  
integration into quantum field theory', {\em Historical Studies in the Physical and  
Biological Sciences} {\bf 21}, pp. 211-235.

Butterfield, J. (2011), `Less is Different: Emergence and Reduction Reconciled', in {\em  
Foundations of Physics} {\bf 41}, 1065-1135. At: Springerlink (DOI  
10.1007/s10701-010-9516-1); http://arxiv.org/abs/1106.0702;
and at: http://philsci-archive.pitt.edu/8355/

Butterfield, J. (2011a), `Emergence, Reduction and Supervenience: a Varied Landscape', {\em  
Foundations of Physics}, {\bf 41}, 920-960. At Springerlink: doi:10.1007/s10701-011-9549-0;  
http://arxiv.org/abs/1106.0704: and at: http://philsci-archive.pitt.edu/5549/

Butterfield, J. and Bouatta N. (2013), `Renormalization for Philosophers', forthcoming in {\em Metaphysics in Contemporary Physics}, a volume of {\em Poznan Studies in Philosophy of Science}, ed. T.Bigaj and C. W\"{u}thrich.

Cao, T.Y. (1993), `New philosophy of renormalization: from the renormalization group to  
effective field theories', in Brown ed. (1993), pp. 87-133.

Cao, T.Y. (1997), {\em Conceptual Developments of Twentieth Century Field Theories},  
Cambridge University Press.

Cao, T.Y., ed. (1999), {\em Conceptual Foundations of Quantum Field Theory}, Cambridge  
University Press.

Cao, T.Y. (1999a), `Renormalization group: an interesting yet puzzling idea', in Cao (ed.)  
(1999), pp. 268-286.


Cao, T.Y and Schweber, S. (1993), `The conceptual foundations and the philosophical aspects  
of renormalization theory, {\em Synthese} {\bf 97}, pp. 33-108.

Cartwright, N. (1983), {\em How the Laws of Physics Lie}, Oxford University Press.


Castellani, E. (2002), `Reductionism, emergence and effective field theories', {\em Studies  
in the History and Philosophy of Modern Physics} {\bf 33}, 251-267.

Dizadji-Bahmani, F., Frigg R. and Hartmann S. (2010), `Who's afraid of Nagelian reduction?',  
{\em Erkenntnis} {\bf 73}, pp. 393-412.

Endicott, R. (1998), `Collapse of the New Wave', {\em Journal of Philosophy} {\bf 95}, pp.  
53-72.

Feynman, R. (1985), {\em QED}, Princeton University Press.

Fodor, J. (1974), `Special Sciences (Or: the disunity of science as a working hypothesis), 
{\em Synthese}  {\bf 28}, pp. 97-115.

Galison, P. (1997), {\em Image and Logic: a material culture of microphysics}, University of  
Chicago Press.

Giere, R. (1995), {\em Science Without Laws}, University of Chicago Press.

Glymour, C. (2013), `Theoretical equivalence and the semantic view of theories', {\em 
Philosophy of Science} {\bf 80}, pp.  286-297.

Halvorson, H. (2013), `The semantic view, if plausible, is syntactic', {\em Philosophy of 
Science} {\bf 80}, pp.  475-478.

Hartmann, S. (2001). `Effective field theories, reductionism and scientific explanation',
{\em Studies in History and Philosophy of Modern Physics} {\bf  32}, pp. 267-304.

Huggett, N. and Weingard, R. (1995), `The  renormalization group and effective field  
theories', {\em Synthese} {\bf 102}, 171-194.

Jaffe, A. (1999), `Where does quantum field theory fit into the big picture?', in Cao (ed.)  
(1999), pp. 136-146.

Jaffe, A. (2008), `Quantum theory and relativity', in {\em Contemporary Mathematics}
(Group Representations, Ergodic Theory, and Mathematical Physics: A Tribute to George W.  
Mackey), R. Doran, C.Moore, and R. Zimmer, (eds.), {\bf 449}, pp. 209–246; available at  
http://arthurjaffe.org

Kadanoff, L. (2009), `More is the same: mean field theory and phase transitions', {\em  
Journal of Statistical Physics} {\bf 137},pp. 777-797.

Kadanoff, L. (2013), `Theories of matter: infinities and renormalization', in {\em The  
Oxford Handbook of the Philosophy of Physics}, ed. R. Batterman, Oxford University Press,  
pp. 141-188.

Kaiser, D. (2005), {\em Drawing Theories Apart: the dispersion of Feynman diagrams in  
postwar physics}, University of Chicago Press.

Kim, J. (1999), `Making sense of emergence', {\em Philosophical Studies} {\bf 95}, pp. 3-36;  
reprinted in Bedau and Humphreys (2008); page reference to the reprint.

Kim, J. (2005), {\em Physicalism, or Something Near Enough},  Princeton University Press.

Kim, J. (2006), `Emergence: Core Ideas and Issues', {\em Synthese} {\bf 151}, pp. 547-559.

Landsman, N. (2007), `Between Classical  and Quantum', in J. Butterfield and J. Earman  
(eds), {\em Handbook of the Philosophy of Physics},  Elsevier: Part A, pp. 417-554. 

Landsman, N. (2013), `Spontaneous Symmetry Breaking in Quantum Systems:
Emergence or Reduction?', forthcoming in  {\em  Studies in the History and Philosophy of 
Modern Physics}; http://arxiv.org/abs/1305.4473

Lautrup, B. and Zinkernagel, H. (1999), `g-2 and the trust in experimental results', {\em  
Studies in the History and Philosophy of Modern Physics} {\bf 30}, pp. 85-110.

Leggett, A (2008), `Realism and the physical world', {\em Reports on Progress in Physics}  
{\bf 71},  022001. 

Marras, A. (2002), `Kim on reduction', {\em Erkenntnis} {\bf 57}, pp. 231-257.

Menon, T. and Callender, C. (2013), `Turn and face the ch-ch-changes: philosophical  
questions raised by phase transitions', in {\em The Oxford Handbook of the Philosophy of  
Physics}, ed. R. Batterman, Oxford University Press, pp. 189-223.

Morrison, M. (2012), `Emergent physics and micro-ontology', {\em Philosophy of Science} {\bf 
79}, pp. 141-166.

Nagel, E. (1961), {\em The Structure of Science: Problems in the Logic of Scientific  
Explanation}, Harcourt.

Nagel, E. (1979), `Issues in the logic of reductive explanations', in his {\em Teleology  
Revisited and other essays in the Philosophy and History of Science}, Columbia University  
Press; reprinted in Bedau and Humphreys (2008); page reference to the reprint.

Needham, P. (2009), `Reduction and emergence: a critique of Kim', {\em Philosophical  
Studies} {\bf 146}, pp. 93-116.

Norton, J. (2012), `Approximation and idealization: why the difference matters', {\em  
Philosophy of Science} {\bf 74}, pp. 207-232.

Norton, J. (2013), `Confusions over reduction and  emergence in the physics of phase  
transitions', available on Norton's website, under `Goodies', at: \\  
http://www.pitt.edu/$\sim$jdnorton/Goodies/reduction$\underline{\;\;}$
emergence/red$\underline{\;\;}$em.html

Psillos, S. (1999), {Scientific Realism: how science tracks truth}, Routledge.

Psillos, S. (2009), {\em Knowing the Structure of Nature}, Palgrave Macmillan.

Putnam, H. (1975), `Philosophy and our mental life', in his collection {\em Mind, Language 
and Reality}, Cambridge University Press, pp. 291-303.
  
Redhead, M. (1999), `Quantum field theory and the philosopher, in Cao (ed.) (1999), pp.  
34-40.


Schaffner, K. (1967), `Approaches to reduction', {\em Philosophy of Science} {\bf 34}, pp.  
137-147.

Schaffner, K. (1976), `Reductionism in biology: problems and prospects' in R. Cohen et al.  
(eds), {\em PSA 1974}, pp. 613-632.

Schaffner, K. (1977), `Reduction, Reductionism, Values, and Progress in the Biomedical
Sciences', in Robert G. Colodny, ed., {\em Logic, Laws, and Life: Some Philosophical 
Complications} Pittsburgh University Press, pp. 143-72.

Schaffner, K. (2006), `Reduction: The Cheshire Cat Problem and a Return to Roots', {\em  
Synthese}, {\bf 151} pp. 377-402.

Schaffner, K. (2013), `Ernest Nagel and reduction', {\em Journal of Philosophy} {\bf 109},  
pp. 534-565.

Schweber, S. (1993), `Changing conceptualization of renormalization theory', in Brown ed.  
(1993), pp. 135-166.

Schweber, S. (1994), {\em QED and the Men who Made It}, Princeton University Press.

Sklar, L. (1967), `Types of intertheoretic reduction', {\em British Journal for the  
Philosophy of Science} {\bf 18}, pp. 109-124.

Sober, E. (1999), `The multiple realizability argument  against reductionism',  {\em  
Philosophy of Science} {\bf 66}, pp. 542-564.

Symanzik, K. (1973), `Infra-red singularities and small-distance behaviour analysis', {\em  
Communications in Mathematical Physics} {\bf 34}, pp. 7-36.

Teller, P. (1989), `Infinite renormalization', {\em Philosophy of Science} {\bf 56}, pp.  
238-257; reprinted with minor corrections as Chapter 7 of his {\em An Interpretive  
Introduction to Quantum Field Theory} (1995), Princeton University Press.


Weinberg, S. (1995), {\em The Quantum Theory of Fields}, volume 1, Cambridge University  
Press.

Weinberg, S. (1995a), {\em The Quantum Theory of Fields}, volume 2, Cambridge University  
Press.

Weinberg, S. (1999), `What is quantum field theory and what did we think it was', in Cao ed.  
(1999), pp. 241-251. Also at: arxiv: hep-th/9702027

Wightman, A. (1999), `The usefulness of a general theory of quantized fields', in Cao (ed.)  
(1999), pp. 41-46.

Wilczek, F. (2005), `Asymptotic freedom: from paradox to paradigm', (Nobel Prize Lecture  
2004), {\em Proceedings of the National Academy of Science}, {\bf 102}, pp. 8403-8413;  
available at: hep-ph/0502113

Wilson, K. (1979), `Problems in Physics with Many Scales of Length', {\em Scientific  
American} {\bf 241}, pp. 158-179.


\end{document}